\documentclass[12pt]{article}
\usepackage{amssymb,latexsym, amsmath,color}
\usepackage{graphicx}
\usepackage{setspace} 

\newtheorem{theorem}{Theorem}
\newtheorem{corollary}[theorem]{Corollary}

\newcommand{\rem}[1]{}
\newcommand{\remfigure}[1]{}
\newcommand{\neu}[2]{{#1}{#2}}

\def\eg{{\it e.g.}\ }  

\newcommand{\Bvec}{\mathbf{B}}

\newcommand{\qvec}{\mathbf{q}}
\newcommand{\pvec}{\mathbf{p}}
\newcommand{\uvec}{\mathbf{u}_s}

\newcommand{\Bsvec}{\mathbf{B}_s}

\newcommand{\vvec}{\mathbf{u}}

\newcommand{\xvec}{\mathbf{x}}

\newcommand{\zvec}{\mathbf{z}}
\newcommand{\zsvec}{{\mathbf{z}_s}}

\newcommand{\curl}{\rm{curl}}
\newcommand{\diver}{\rm{div}}
\newcommand{\lansa}{{\sl LANS$-\alpha$ }}
\newcommand{\lamhda}{{\sl LAMHD$-\alpha$ }}
\newcommand{\zerovec}{\mathbf{0}}
\newcommand{\half}{\frac{1}{2}}
\renewcommand{\vec}{\mathbf}
\begin{document}


\title{
Inertial Range Scaling, K\'arm\'an-Howarth Theorem and 
Intermittency for Forced and Decaying Lagrangian Averaged MHD in 2D}
\author{J. Pietarila Graham$^{1}$, D. D. Holm$^{2,3}$, P. Mininni$^{1}$ and A. Pouquet$^{1}$
\rem{
\thanks{
Darryl D. Holm,
Computer and Computational Science Division, Los
Alamos National Laboratory, MS D413, Los Alamos, NM 87545. 
email: dholm@lanl.gov, phone: (505) 667-6398, fax: (505) 665-5757,
http://math.lanl.gov/~dholm/ and
Mathematics Department, Imperial College London, SW7 2AZ, UK. email:
d.holm@imperial.ac.uk, 
phone: +44 (0)20 7594 8531,
fax: +44 (0)20 7594 8517,
http://www.ma.ic.ac.uk/~dholm/
}
}
}
\date{\small(Dated: Mar 1, 2006)}

\maketitle

\begin{abstract}
We present an extension of the K\'arm\'an-Howarth theorem to the Lagrangian 
averaged magnetohydrodynamic ({\sl LAMHD$-\alpha$}) equations. The scaling laws resulting 
as a corollary of this theorem are studied in numerical simulations, as 
well as the scaling of the longitudinal structure function exponents 
indicative of intermittency. Numerical simulations
for a magnetic Prandtl number equal to unity
 are presented both for 
freely decaying and for forced two dimensional MHD turbulence, solving 
directly the MHD equations, and employing the \lamhda equations at 1/2 and 
1/4 resolution. Linear scaling of the third-order structure function with 
length is observed. The \lamhda equations also capture the anomalous scaling 
of the longitudinal structure function exponents up to order 8.
\end{abstract}
$^1$ National Center for Atmospheric Research, P.O. Box 3000, Boulder, CO  80307,  USA \\
$^2$ Department of Mathematics, Imperial College London, London SW7 2AZ, UK \\
$^3$ Computer and Computational Science Division, Los Alamos National Laboratory, Los Alamos, NM  87545, USA
\newpage

\section{Introduction}
Turbulent flows are ubiquitous in Nature and their very complexity make 
them hard to understand, let alone predict. One reason is that turbulence 
comes in intermittent ``gusts'', such as fronts. These gusts are
associated with \neu{}{their} non-Gaussian statistics, which allows strong events
occurring in the ``fat tails'' of the statistical distribution of, say, 
velocity gradients and other physical variables. From the
theoretical viewpoint, its property of intermittency is the {\it sine qua
non} of turbulence. Intermittency may also trigger large scale effects,
such as the apparently random reversals of the Earth's magnetic field
\neu{}{\cite{HOS01}.  It plays a role in other large scale events in the 
solar dynamo, and in the atmosphere of the earth. Although intermittency 
is believed to take place 
at small scales, strong events can affect the dynamics of the large 
scales, specially in systems close 
to criticality. For instance, Ref. \cite{H93} shows that local fluctuations 
of the kinetic helicity can explain phase and amplitude variations of
the 22-years solar cycle. Intermittency is also one possible explanation 
for the occurrence of the Maunder-like minima of solar activity 
\cite{CBLSJ04}. 
Finally, intermittency is known to affect the transport of momentum in 
atmospheric surface layers \cite{KSM99}.}

Intermittency is a highly spatially and temporally localized phenomenon, 
which thus requires high-resolution instrumentation, be it in the 
laboratory, in atmospheric and geophysical flows or in numerical 
simulations. In the latter case, the lack of adequate computer power 
(from the standpoint of the geo-physicist) implies that modeling of 
the unresolved small scales must take place. However, intermittency in general is 
not included explicitly in Large Eddy Simulations (LES) models of
turbulent processes.

The intermittency of the subgrid-scale (SGS) energy dissipation as is 
usually defined in LES has been studied filtering 3D direct numerical 
simulations (DNS) of non-conductive fluids \cite{CM98}, with a Taylor 
Reynolds number $R_\lambda=150$; such a Reynolds number is somewhat insufficient 
for a determination of third-order scaling. For the SGS dissipation, they
find that the Smagorinsky model, the volume-averaged dynamic model, and the
similarity model perform fairly well ({\sl e.g.} the error in the exponent 
for $p=7$ is less than 7\%). On the other hand, the constant eddy-viscosity 
and spectral eddy-viscosity models underestimate intermittency beyond 
$p=4$ (compared to DNS) while the local and clipped dynamic Smagorinsky 
models strongly overestimate the intermittency beyond $p=4$. In models 
not capturing intermittency properly, the question arises whether the overall
statistics of the flow at large scales would be affected by the absence of
intermittency; and if so, how intermittency should be incorporated. 

There are many models of turbulent flows (see for example the recent 
review in \cite{meneveaukatz}) and there are many turbulent flows for 
which adequate testing of such models is in order. Magnetohydrodynamics, 
{\it i.e.} the coupling of a velocity and magnetic field at sub-luminal velocities so that the
displacement current can be neglected, presents an 
interesting property, namely that in two space dimensions (2D), 
intermittency occurs as well as in three dimensions (3D).
This is in contrast with the 2D neutral fluid 
case for which the conservation of vorticity leads to an inverse energy 
cascade to the large scales; in the presence of a magnetic field, this 
conservation is  broken by the Lorentz force. Since, from a numerical 
standpoint, much higher Reynolds numbers can be achieved in 2D, an 
intermittent flow can be reached in 2D-MHD with adequate scale separation between 
the energy-containing range, the inertial range and the dissipation range. 
Our 2D MHD tests are able to exhibit a substantially larger Reynolds number 
(up to $R_\lambda\sim 1500$) than the values listed for previous studies. 
This fact provides an ideal testing ground for models of turbulent MHD 
flows, such as they occur in geophysics and astrophysics: magnetic fields 
are observed in detail in the Earth and Sun environments, and are known to 
be dynamically important as well for the solar-terrestrial interactions 
(the so-called space weather), in the interstellar medium and in galaxies.

Modeling of MHD flows is still under development (see \cite{LESMHD}). 
\neu{}{Most LES for hydrodynamic turbulence are based upon self-similarity 
or universality, in that they assume a known power law of the energy spectrum.
For MHD, the kinetic energy is not a conserved quantity, and this poses a
problem for the extension of such techniques to the case of MHD.
Additional difficulties arise from the fact that MHD has several regimes
depending on the relative strengths of the magnetic and velocity fields,
their degree of alignment, and whether mechanical or magnetic energy is
injected into the flow.  However, some LES have been developed for particular
cases.  There exists LES for MHD turbulence with some degree of alignment 
between the velocity and magnetic fields \cite{AMK+01},
dissipative LES which does not model the interactions between the two fields
\cite{TFS94}, and LES for low magnetic Reynolds number \cite{KM04}.
We have recently tested one model which may be more generally applicable,}
the Lagrangian averaged
magneto-hydrodynamics alpha ({\sl LAMHD$-\alpha$}) model, both in 2D 
\cite{lamhd2d} and in 3D \cite{lamhd3d} and it has been used to examine 
the onset of the dynamo instability when the magnetic Prandtl number (the 
ratio of viscosity $\nu$ to magnetic diffusivity $\eta$) is small 
\cite{prldynamo}, as occurs in liquid metals in the laboratory, in the 
liquid core of the Earth or in the solar convection zone.
In this context, because of the importance of intermittency as a 
fundamental, or even defining property of turbulence, we seek to 
determine to what extent \lamhda exhibits intermittency.

Intermittency is believed to be associated only with a forward cascade of
energy; that is, the cascade of energy from larger scales to smaller
scales, or, equivalently, from low wave numbers to high wave numbers. As 
previously mentioned, in determining the extent to which the 
{\sl LAMHD$-\alpha$} model exhibits intermittency, we shall take 
advantage of the forward cascade of energy which occurs in 
two-dimensional MHD. We shall first investigate the spectral 
scaling laws for this situation in Sec. \ref{scaling-sec},
then its K\'arm\'an-Howarth theorem in Sec. \ref{KH-sec}. Section
\ref{KH-sec} also discusses the modifications of the K\'arm\'an-Howarth
theorem for MHD which arise due to the presence of the length scale alpha
($\alpha$) in the {\sl LAMHD$-\alpha$} model. The length scale alpha
modifies the nonlinearity in the motion equation, and one must
estimate its observable physical effects. In particular, introduction of
the length scale alpha modifies the {\sl LAMHD$-\alpha$} energy spectrum
for $k\alpha>1$. Section \ref{erg-anom-sec} discusses energy conservation 
for the {\sl LAMHD$-\alpha$} model and investigates its inviscid energy 
dissipation anomaly, which arises from its scaling laws and its 
K\'arm\'an-Howarth theorem. Finally in Sec. \ref{scale-anom-sec} we 
investigate the effects of introducing the length scale alpha on the 
intermittency of the {\sl LAMHD$-\alpha$} model solutions for decaying 
and forced turbulence in two dimensions. Just as for Navier-Stokes 
turbulence, these effects emerge in numerical simulations as a scaling 
anomaly in the higher order structure functions of the 
{\sl LAMHD$-\alpha$} model. Section \ref{discuss-sec} summarizes
these results.

\section{Background of the {\sl LAMHD$-\alpha$} model}
\subsection{The {\sl LAMHD$-\alpha$} model equations}
The Lagrangian-averaged magnetohydrodynamic alpha, or \lamhda model was
derived by Lagrangian averaging ordinary MHD along particle trajectories
\cite{Ho2002}.  Specifically, the \lamhda equations arise from 
Lagrangian-averaging Hamilton's principle for incompressible ideal MHD,
after using a form of Taylor's hypothesis of frozen-in turbulent
fluctuations in the Euler-Poincar\'e equation for barotropic MHD from
Ref.~\cite{HoMaRa1998b}. When Navier-Stokes viscosity $\nu$ and
diffusivity $\eta$ are included in the standard fashion, the
equations for the \lamhda model emerge as,
\begin{eqnarray}
&\partial_t \vvec + \uvec\cdot\nabla\vvec - \Bsvec\cdot\nabla\Bvec
  + (\nabla\uvec)^T\cdot\vvec + (\nabla\Bvec)^T\cdot\Bsvec  
  + \nabla\pi = \nu\Delta\vvec, \label{eq:lamhd1} \\
&\partial_t \Bsvec + \uvec\cdot\nabla\Bsvec - \Bsvec\cdot\nabla\uvec 
   =\eta\Delta\Bvec , \label{eq:lamhd2} \\
& \diver\,\uvec = 0, 
  \qquad \diver\,\Bsvec = 0
\,.\label{eq:lamhd3}
\end{eqnarray}
In these equations, subscript $s$ denotes the smoothing obtained by
inverting the Helmholtz relations,
\begin{eqnarray}
\vvec = (1-\alpha^2\Delta)\uvec, \quad 
\Bvec = (1-\alpha_M^2\Delta)\Bsvec 
\,.
\label{eq:helmholtz}
\end{eqnarray}
with Dirichlet boundary conditions,
\[\uvec=0\quad\hbox{and}\quad\Bsvec=0\quad\hbox{on the boundary}\,.\]
The modified total pressure $\pi$ in the motion equation for the \lamhda
model is defined by
\begin{eqnarray}
\pi + \half|\Bsvec|^2 
= 
p - \half|\uvec|^2 - \frac{\alpha^2}{2}|\nabla\uvec|^2   
\,,\label{pi-def}
\end{eqnarray}
where $p$ is the mechanical pressure. In these equations, $\alpha$ and
$\alpha_M$ are two constant parameters: $\alpha$ characterizes the 
correlation length between the instantaneous Lagrangian fluid trajectory
and its mean (time average); while $\alpha_M$ is its magnetic
counterpart. These two parameters need not be equal, {\it ab initio}.  The
traditional MHD system is obtained by setting both $\alpha = 0$ and
$\alpha_M=0$.  Likewise, the \lansa incompressible fluid turbulence model
is obtained by setting $\Bsvec = \zerovec$. These equations may also be
obtained through a filtering approach, as proposed in the fluid case in
\cite{montgo_02}.

\subsection{Energy, momentum, circulation and linkages}
The \lamhda system of equations (\ref{eq:lamhd1}-\ref{eq:lamhd3}) possesses
the standard properties of a normal fluid theory.  For
example, the
\lamhda system monotonically dissipates the positive energy
\begin{equation}\label{erg-bar-incomp-MHD-alpha}
{\cal E}
=
\rem{
\frac{1}{2}
\int \Big[\
|\mathbf{u}_s|^2
+
\alpha^2
|\nabla\mathbf{u}_s|^2
+
|\mathbf{B}_s|^2
+
\alpha^2
|\nabla\mathbf{B}_s|^2
\Big]\,d^3x
=}
\frac{1}{2}
\int 
\Big(\mathbf{u}_s\cdot\mathbf{u}
+
\mathbf{B}_s\cdot\mathbf{B}
\Big)\,d^3x
\,,
\end{equation}
according to
\begin{equation}\label{erg-dis}
\frac{d{\cal E}}{dt}
=
-\,\nu
\int 
\Big(|\nabla\mathbf{u}_s|^2
+
\alpha^2
|\Delta\mathbf{u}_s|^2
\Big)\,d^3x
-\,\eta
\int 
\Big(|\nabla\mathbf{B}|^2
+
\alpha^2
|\Delta\mathbf{B}|^2
\Big)\,d^3x
\,.
\end{equation}
In addition, the \lamhda motion equation (\ref{eq:lamhd1}) may be expressed in
conservative form as
\begin{equation}\label{mom-cons}
\frac{\partial }{\partial  t}\,u_i
= 
-\ \frac{\partial }{\partial  x^j}
\Big(T\,^j_i-\nu\, u_{i,l}\delta^{lj}\Big)
\,,
\end{equation}
with stress tensor
\begin{eqnarray} \label{stress-tens}
T\,^j_i 
&=&
\Big(u_iu_s^j - \alpha^2 u_{s\,k,i} u_s^{\,k,j}\Big)
-\Big(
B_iB_s^j - \alpha^2 B_{s\,k,i} B_s^{\,k,j}
\Big)
\nonumber \\
&&
+\
\delta^j_i
\Big(
p - \frac{1}{2}|\mathbf{B}_s|^2 + \mathbf{B}_s\cdot\mathbf{B}
- \frac{\alpha^2}{2} |\nabla\mathbf{B}_s|^2
\Big)
\,.
\end{eqnarray}
Thus, the two velocities appearing in the \lamhda model may be interpreted as
fluid parcel velocity $\mathbf{u}_s$ and momentum per unit mass $\mathbf{u}$. 

The Kelvin circulation theorem for the incompressible
\lamhda motion equation (\ref{eq:lamhd1}) involves both of these velocities,
\begin{equation}\label{kel-circ-lamhd-alpha}
\frac{d}{dt}\oint_{c(\mathbf{u}_s)}\!\!\!
\mathbf{u}
\cdot
d\mathbf{x}
=
\oint_{c(\mathbf{u}_s)}\!\!\!
(\mathbf{J}\times\mathbf{B}_s + \nu\Delta\mathbf{u})
\cdot
d\mathbf{x}
\,,
\end{equation}
where $\mathbf{J}
=
\,{\rm curl}\,\mathbf{B}$.
Hence, the $\mathbf{J}\times\mathbf{B}_s$ force and
viscous force can each generate circulation of $\mathbf{u}$ around material loops
moving with smoothed velocity $\mathbf{u}_s$. This results by Stokes theorem in
vorticity dynamics for $\boldsymbol{\omega}={\rm curl}\,\mathbf{u}$ in the form
\begin{equation}\label{vortex-dyn-lamhd-alpha}
\frac{\partial \boldsymbol{\omega}}{\partial  t}
+
\mathbf{u}_s\cdot\nabla\boldsymbol{\omega}
-\boldsymbol{\omega}\cdot\nabla\mathbf{u}_s
=
\mathbf{B}_s\cdot\nabla\mathbf{J}
-\mathbf{J}\cdot\nabla\mathbf{B}_s
+
\nu\Delta\boldsymbol{\omega}
\,.
\end{equation}
The linkages of the smooth B-field $\mathbf{B}_s$ with itself and with the
vorticity $\boldsymbol{\omega}$ are given respectively by the helicity 
$\int\mathbf{A}_s\cdot\mathbf{B}_s\,d\,^3x$ and cross helicity
$\int\mathbf{u}\cdot\mathbf{B}_s\,d\,^3x$. The densities for these linkages 
satisfy
\begin{equation}\label{hel-dyn-lamhd-alpha}
\frac{\partial }{\partial  t}(\mathbf{A}_s\cdot\mathbf{B}_s)
+
\,{\rm div}\,\big((\mathbf{A}_s\cdot\mathbf{B}_s)\,\mathbf{u}_s\big)
=
\eta\,(\mathbf{A}\cdot\mathbf{B}_s+\mathbf{A}_s\cdot\mathbf{B})
\,,
\end{equation}
in which $\mathbf{B}_s={\rm curl}\,\mathbf{A}_s$ and $\mathbf{B}={\rm
curl}\,\mathbf{A}$, and, cf. (\ref{pi-def}),
\begin{equation}\label{crosshel-dyn-lamhd-alpha}
\frac{\partial }{\partial  t}(\mathbf{u}\cdot\mathbf{B}_s)
+
\,{\rm div}\,\big((\mathbf{u}\cdot\mathbf{B}_s)\,\mathbf{u}_s
+(\pi+\frac{1}{2}|\mathbf{B}_s|^2)\mathbf{B}_s\big)
=
\nu\,\mathbf{B}_s\cdot\Delta\mathbf{u}
+\eta\,\mathbf{u}\cdot\Delta\mathbf{B}
\,.
\end{equation}
Thus, resistivity affects the helicity, while both resistivity and viscosity
affect the cross helicity, and these linkages are both preserved by \lamhda in
the ideal case. Of course, these properties of energy, momentum, circulation and
linkages for the \lamhda model all reduce to characteristics of normal MHD, when
$\alpha^2\to0$.

\subsection{Recasting \lamhda as an LES turbulence model}
The \lamhda model modifies the motion equation for ordinary MHD.  By a
short sequence of manipulations, we may recast the \lamhda motion equation
into a form which is reminiscent of an LES turbulence model. We begin with
the following commutation relation,
\begin{eqnarray}
\left [ \pvec\cdot\nabla,(1-\alpha^2\Delta) \right ]\qvec
 = \alpha^2\diver\,\left ( \nabla\qvec\cdot\nabla\pvec +
 \nabla\qvec\cdot\nabla\pvec^T \right ) 
- \alpha^2\left ( \nabla(\diver\,\pvec)\cdot\nabla\right )\qvec
\label{eq:id0}
\end{eqnarray}
which holds for any vectors $\pvec$ and $\qvec$. Two other useful
vector identities are,
\begin{eqnarray}
&(\nabla\uvec)^T\cdot\vvec - \nabla\left ( \half|\uvec|^2 +
\frac{\alpha^2}{2}|\nabla\uvec|^2  \right ) =
-\alpha^2\diver\,(\nabla\uvec^T\cdot \nabla\uvec), \label{eq:id1} \\
&(\nabla\Bvec)^T\cdot\Bsvec - \nabla \left ( \half \left (
1-\alpha^2\Delta\right )|\Bsvec|^2    
  + \frac{\alpha^2}{2}|\nabla\Bsvec|^2 \right ) = 
  \alpha^2\diver\,(\nabla\Bsvec^T\cdot \nabla\Bsvec) \label{eq:id2} 
\hspace{5mm}
\end{eqnarray}
where $\alpha = \alpha_M$ was assumed. Consequently, the motion 
equation in the incompressible \lamhda model may be rewritten 
equivalently in ``LES form'' as,
\neu{}{
\begin{eqnarray}
(1-\alpha^2\Delta)\left ( 
  \partial_t\uvec + \uvec\cdot\nabla\uvec - \Bsvec\cdot\nabla\Bsvec 
  + \nabla p_s - \nu\Delta\uvec \right )
  = - \alpha^2\diver\,\tau,
\label{eq:lamhdU}
\end{eqnarray}
}
where the divergence of the ``stress tensor'' $\tau$ is given by
\begin{eqnarray}
\diver\,\tau &=& \diver\,(\nabla\uvec\cdot\nabla\uvec +
 \nabla\uvec\cdot\nabla\uvec^T - \nabla\uvec^T\cdot \nabla\uvec)
  \nonumber\\&&- \,\diver\,(\nabla\Bsvec\cdot\nabla\Bsvec +
 \nabla\Bsvec\cdot\nabla\Bsvec^T - \nabla\Bsvec^T\cdot \nabla\Bsvec) 
\,,
\label{def:div-tau}
\end{eqnarray}
and gradient terms have been absorbed into the modified total
pressure, \neu{}{denoted by $\tilde\pi$}, which is given by
\begin{eqnarray}
\tilde{\pi} = p - \half|\uvec|^2 - \frac{\alpha^2}{2}|\nabla\uvec|^2  
  - \half \left ( 1-\alpha^2\Delta\right )|\Bsvec|^2   
  - \frac{\alpha^2}{2}|\nabla\Bsvec|^2,
\end{eqnarray}
\neu{}{where $p = (1-\alpha^2\Delta)p_s.$}
By using the following identity for divergenceless vectors 
${\rm\,div\,}\mathbf{u}=0$
\begin{equation} \label{identity}{\rm\,div\,}\big(\nabla\mathbf{u}^T
  \cdot\nabla\mathbf{u}^T\,\big)=\nabla \frac{1}{2}{\rm\,tr\,}
  (\nabla\mathbf{u}\cdot\nabla\mathbf{u})\,,\end{equation}
we may rewrite the added stress in (\ref{def:div-tau})
equivalently, as
\begin{eqnarray}
\diver\,\tau &=& 4\,\diver\,(S\cdot\Omega - \Sigma\cdot J)+ \nabla \Pi
\end{eqnarray}
with new notation
\begin{eqnarray*}
S &=& \frac{1}{2}\big( \nabla\uvec + \nabla\uvec^T\,\big)
\,,\quad
\Omega = \frac{1}{2}\big( \nabla\uvec - \nabla\uvec^T\,\big)
\\
\Sigma &=& \frac{1}{2}\big( \nabla\Bsvec + \nabla\Bsvec^T\,\big)
\,,\quad
J = \frac{1}{2}\big( \nabla\Bsvec - \nabla\Bsvec^T\,\big)
\,,
\end{eqnarray*}
and additional pressure
\begin{eqnarray*}
\Pi =  \frac{1}{2}{\rm\,tr\,}(\nabla\uvec\cdot\nabla\uvec)
-
 \frac{1}{2}{\rm\,tr\,}(\nabla\Bsvec\cdot\nabla\Bsvec)
\,.
\end{eqnarray*}
The (non-symmetric) stress tensor  $\,\tau$ given by
\begin{eqnarray}
\tau = 4\,(S\cdot\Omega - \Sigma\cdot J)+ {\rm Id}\, \Pi
\end{eqnarray} 
which emerges from these manipulations casts the \lamhda model into a
form reminiscent of an LES turbulence model. We shall find these expressions 
convenient below in introducing the analog of Els\"asser variables for the
\lamhda model.

\section{Inertial range scaling laws in forward energy cascade}
\label{scaling-sec}

In two space dimensions, the conservation of vorticity in the neutral 
(${\bf B} \equiv 0$ case) leads to an inverse energy cascade; however, 
the Lorentz force breaks this conservation and, in MHD, energy is 
found to be mostly transferred to the small scales both in 2D and in 3D.
Several measurements, starting with satellite data in the solar wind and 
continuing more recently with direct numerical simulations both in two 
dimensions and three dimensions, indicate that the energy spectrum of a 
turbulent MHD flow follows a law that is barely distinguishable from a 
neutral fluid, with $E(k)\sim k^{-1.70}$. Differences do occur when one 
examines higher order structure functions: the most intermittent case 
(almost comparable in magnitude to that of the passive scalar) is the 
two-dimensional MHD fluid; three-dimensional MHD appears less intermittent 
than the 2D MHD case \cite{biskamp}, and the 3D neutral fluid is the lesser intermittent 
of the three. Intermittency has been observed in the Solar Wind as well.
None of the data at the second order level is in agreement with the 
phenomenologies developed by Kolmogorov \cite{K41b} and leading to an 
energy spectrum $E(k)\sim k^{-5/3}$ (heretofore the K41 model) or by 
Iroshnikov and by Kraichnan \cite{iroshnikovRHK} (heretofore, the 
IK model) and leading to a shallower spectrum, {\it viz.} 
$E(K)\sim k^{-3/2}$ in the absence of significant velocity-magnetic field 
correlations. These two types of phenomenology differ by the taking into 
account in the latter case of the non-local interactions (in Fourier space) 
emanating from the propagation of Alfv\'en waves; it is worth mentioning 
here that the IK model also agrees with the isotropic limit of the weak 
turbulence theory for incompressible MHD \cite{galtier2000}. Note also 
that a model of intermittency for MHD flows \cite{PoPu1995} does recover 
the intermittency as measured in direct numerical simulations both in 
2D \cite{PoPu1995} and in 3D \cite{biskamp}, but such models depend 
on two adjustable parameters and thus do not necessarily have a 
predictive power.

These anomalous scaling laws are not fully understood but, for the neutral 
case, there is an exact law at third order with which the K41 phenomenology 
is compatible. In MHD, the exact law is more complex in its structure since 
it involves third-order cross-correlations between the velocity and the 
magnetic field \cite{PoPu1998} whereas the phenomenologies evoked above 
refer to single-variable moments. In that instance, it is worth asking 
what is the equivalent exact law in the context of the \lamhda model, a 
task developed in the next section. The question also arises as to what 
is the spectrum of energy beyond the alpha cut-off scales (which we take 
equal here, although different choices can be made, see {\it e.g.} 
\cite{lamhd3d,prldynamo}). The answer should be guided by what is the 
pseudo-invariant in the small scales, beyond $\alpha$. In the neutral 
fluid case, it is the enstrophy $\left<\omega^2\right>$ and in MHD this 
becomes $\left<\omega^2\right>+\left<j^2\right>$, where ${\bf j}$ is 
the current density. A Kolmogorov-like dimensional reasoning (see 
\cite{FoHoTi2001} for the \lansa case) taking into account this pseudo-invariance law
will lead to a $k^{-3}$ spectrum 
at scales smaller than $\alpha$ whereas it can be easily shown that
the corresponding IK arguments 
lead to a $k^{-5/2}$ law.

\section{K\'arm\'an-Howarth theorem for \lamhda in 2D and 3D
Els\"asser variables} \label{KH-sec}

In 1938, K\'arm\'an and Howarth \cite{KaHo1938} introduced the invariant 
theory of isotropic hydrodynamic turbulence, and derived from the 
Navier-Stokes equations an exact law relating the time derivative of the 
two-point velocity correlation with the divergence of the third-order 
correlation function. Later, this result was generalized to the MHD 
case by Chandrasekar \cite{Ch1951}, and recently written in terms of 
Els\"asser variables \cite{PoPu1998}. For \lansa in the fluid case it was derived in 
\cite{Ho2002c}. The relevance of the K\'arm\'an-Howarth theorem for 
the study of turbulence cannot be underestimated. As a corollary, rigorous 
scaling laws in the inertial range can be deduced. In this section we 
will generalize these results to the \lamhda case.

For the sake of simplicity, we will consider the case 
$\eta = \nu = 0$, the dissipative terms can be added at any point 
in the derivation. Also, we will use $\alpha=\alpha_M$. We start 
writing the {\sl LAMHD$-\alpha$} equations using the Els\"asser 
variables
\begin{equation}
\zvec^\pm = \vvec \pm \Bvec , \quad
\zsvec^\pm = \uvec \pm \Bsvec .
\label{eq:elsasser}
\end{equation}

Applying the Helmholtz operator to eq. \neu{}{(\ref{eq:lamhd2})}, we obtain
\begin{equation}
\left( 1-\alpha^2 \Delta \right) \left( \partial_t \Bsvec + 
    \uvec\cdot\nabla\Bsvec - \Bsvec\cdot\nabla\uvec \right) = 0.
\label{eq:lamhdB}
\end{equation} 
Now we add and subtract eqs. (\ref{eq:lamhdU}) and (\ref{eq:lamhdB}). 
Using eqs. (\ref{eq:elsasser}) we obtain equations for the evolution of 
$\zsvec^\pm$,
\begin{equation}
\left( 1- \alpha^2 \Delta \right) \left( \partial_t \zsvec^\pm + \zsvec^\mp 
    \cdot \nabla \zsvec^\pm + \nabla \tilde{\pi}_s \right) = 
    - \alpha^2 \diver\, \tau , 
\label{eq:Elsasser1}
\end{equation}
where the stress tensor divergence $\diver\,\tau$ in terms of the Els\"asser 
variables is
%
\begin{eqnarray}
\diver\,\tau &=& \frac{1}{2}\, \diver\left( \nabla \zsvec^+ \cdot \nabla 
    \zsvec^- + 
    \nabla \zsvec^+ \cdot {\nabla \zsvec^-}^T - {\nabla \zsvec^+}^T 
    \cdot \nabla \zsvec^- \right. \nonumber \\
&& \left. + \nabla \zsvec^- \cdot \nabla \zsvec^+ + \nabla \zsvec^- 
    \cdot {\nabla \zsvec^+}^T - {\nabla \zsvec^-}^T \cdot \nabla 
    \zsvec^+ \right) .
\end{eqnarray}
This stress divergence may be rewritten equivalently, as
\begin{eqnarray}
\diver\,\tau &=& 2\,\diver\,\Big(\Delta^+\cdot\Sigma^- 
+ \Delta^-\cdot\Sigma^+\Big)
+ \nabla \Pi
  \nonumber\\
\hbox{with}\!\!\!\!&&\!\!\!\!
\Sigma^\pm = \frac{1}{2}\Big( \nabla\zvec^\pm +( \nabla\zvec^\pm)^T\,\Big)
\,,\quad
\Delta^\pm = \frac{1}{2}\Big( \nabla\zvec^\pm -( \nabla\zvec^\pm)^T\,\Big)
  \nonumber\,,
\end{eqnarray}
with the same additional pressure $\Pi$ as before.

We could repeat all the derivation to obtain an equation for the evolution 
of $\zvec^\pm$ from eqs. (\ref{eq:lamhd1}) and (\ref{eq:lamhdB}). Instead, 
starting from eq. (\ref{eq:Elsasser1}), using eqs. (\ref{eq:helmholtz}) and 
(\ref{eq:id0}) we obtain
\begin{equation}
\partial_t \zvec^\pm + \zsvec^\mp  \cdot \nabla \zvec^\pm + \nabla 
    \tilde{\pi} = \alpha^2 \diver\, \varsigma^\pm \,
\label{eq:Elsasser2}
\end{equation}
where
\begin{eqnarray}
\varsigma^\pm &=& \frac{1}{2} \left( \nabla \zsvec^\pm \cdot \nabla 
    \zsvec^\mp + \nabla \zsvec^\pm \cdot {\nabla \zsvec^\mp}^T + 
    {\nabla \zsvec^\pm}^T \cdot \nabla \zsvec^\mp  \right. \nonumber \\
&& \left. - \nabla \zsvec^\mp \cdot \nabla \zsvec^\pm - \nabla \zsvec^\mp 
    \cdot {\nabla \zsvec^\pm}^T  + {\nabla \zsvec^\mp}^T \cdot \nabla 
    \zsvec^\pm \right) .
\end{eqnarray}

Note that equations (\ref{eq:Elsasser1}) and (\ref{eq:Elsasser2}) make 
explicit the fact that Alfv\'en waves 
$\vvec = \pm \Bvec$, $\uvec = \pm \Bsvec$ are exact nonlinear 
solutions of the \lamhda equations.
\neu{}{For an  Alfv\'en wave either $\zvec^+$ or $\zvec^-$ (as well as 
the corresponding field $\zsvec^\pm$) is
zero.  In this case, all nonlinear terms are zero and verification of the
solution follows.}

In Cartesian coordinates, we can write equations (\ref{eq:Elsasser1}) and 
(\ref{eq:Elsasser2}) in components
\begin{eqnarray}
& \partial_t z_i^\pm + \partial_k \left( z_i^\pm {z_s}^{\mp k} + \tilde{\pi} 
    \delta_i^k - \alpha^2 {\varsigma^\pm}_i^k \right) = 0 \label{eq:comp1} \\
& \partial_t {z'_s}_j^\pm + {\partial'}_k \left( {z'_j}^\pm {z'_s}^{\mp k} + 
    {\tilde{\pi}'}_s \delta_j^k + \alpha^2 g_\alpha * {\tau '}_j^k \right) 
    = 0 , \label{eq:comp2}
\end{eqnarray}
the prime denotes that the variables are evaluated at $\xvec'$, and
\begin{equation}
g_\alpha = \frac{e^{-r/\alpha}}{4 \pi \alpha^2 r}
\end{equation}
is the Yukawa potential. The Green function of the Helmholtz operator 
is given by
\begin{equation}
g_\alpha * \tau_i^k = \int{ g_\alpha \left( |\xvec ' - \xvec| \right) 
    \tau_i^k (\xvec ') d^3 x'} ,
\end{equation}
and the components of the stress tensors $\tau$ and $\varsigma^\pm$ are
\begin{eqnarray}
\tau_i^k &=& \frac{1}{2} \left( \partial_j {z_s}_i^+ 
    \partial^k {{z_s}^-}^j + \partial_j {z_s}_i^+ \partial^j {{z_s}^-}^k 
    - \partial_i {z_s}_j^+ \partial^k {{z_s}^-}^j \right. \nonumber \\
{} && \left. + \partial_j {z_s}_i^- \partial^k {{z_s}^+}^j + 
    \partial_j {z_s}_i^- \partial^j {{z_s}^+}^k - \partial_i {z_s}_j^- 
    \partial^k {{z_s}^+}^j \right) , \\
{\varsigma^\pm}_i^k &=& \frac{1}{2} \left( \partial_j {z_s}_i^\pm 
    \partial^k {{z_s}^\mp}^j + \partial_j {z_s}_i^\pm \partial^j {{z_s}^\mp}^k 
    + \partial_i {z_s}_j^\pm \partial^k {{z_s}^\mp}^j \right. \nonumber \\
{} && \left. - \partial_j {z_s}_i^\mp \partial^k {{z_s}^\pm}^j - 
    \partial_j {z_s}_i^\mp \partial^j {{z_s}^\pm}^k + \partial_i {z_s}_j^\mp 
    \partial^k {{z_s}^\pm}^j \right) .
\end{eqnarray}

Multiplying eq. (\ref{eq:comp1}) by ${z'_s}_j^\pm$, eq. (\ref{eq:comp2}) by 
$z_i^\pm$, and adding the result yields
\begin{eqnarray}
\partial_t \left<z_i^\pm {z'_s}_j^\pm \right> &=& \frac{\partial}{\partial r^k}
    \left< \left( z_i^\pm {z_s}^{\mp k} - \alpha^2 {\varsigma^\pm}_i^k \right) 
    {z'_s}_j^\pm \right> + \frac{\partial}{\partial r^k} \left< \tilde{\pi} 
    {z'_s}_j^\pm \delta_i^k - \tilde{\pi}_s' z_i^\pm \delta_j^k \right> 
\nonumber \\
    {} & & - \frac{\partial}{\partial r^k} \left< \left( {z'_s}_j^\pm 
    {z'_s}^{\pm k} + \alpha^2 g_\alpha * {\tau'}_j^k \right) z_i^\pm \right> ,
\label{iso-homo-stat}
\end{eqnarray}
where we used homogeneity
\begin{equation}
\frac{\partial}{\partial r^k} \left< \cdot \right> = 
    \frac{\partial}{\partial x'^k} \left< \cdot \right> = 
    - \frac{\partial}{\partial x^k} \left< \cdot \right> .
\end{equation}

Now, we can make the equation symmetric in the indices $i,j$ adding 
the equation for $\partial_t \left< z_j^\pm {z'_s}_i^\mp \right>$. 
We use homogeneity 
\begin{equation}
\left< q_i {q'_s}_j {q'_s}^k + q_j {q'_s}_i {q'_s}^k \right> = 
    - \left< q'_i {q_s}_j {q_s}^k + q'_j {q_s}_i {q_s}^k \right>,
\end{equation}
and define the tensors
\begin{eqnarray}
{\cal Q}_{ij}^\pm &=& \left< z_i^\pm {z'_s}_j^\pm + z_j^\pm {z'_s}_i^\pm  
    \right> , \\
{{\cal T}^\pm}_{ij}^k &=& \left< \left(z_i^\pm {z'_s}_j^\pm + 
    z_j^\pm {z'_s}_i^\pm + {z'_i}^\pm {z_s}_j^\pm +
    {z'_j}^\pm {z_s}_i^\pm \right) {z_s}^{\mp k} \right> , \label{eqT} \\
{\Pi^\pm}_{ij}^k &=& \left< \left( \tilde{\pi}_s' z_j^\pm - \tilde{\pi} 
    {z'_s}_j^\pm \right) \delta_i^k + \left( \tilde{\pi}_s' z_i^\pm - 
    \tilde{\pi} {z'_s}_i^\pm \right) \delta_j^k \right> , \\
{{\cal S}^\pm}_{ij}^k &=& \left< \tau_i^k {z'_s}_j^\pm + \tau_j^k {z'_s}_i^\pm 
    + g_\alpha * {\tau'_s}_j^k z_i^\pm + g_\alpha * {\tau'_s}_i^k 
    z_j^\pm \right> . \label{eqS}
\end{eqnarray}
We can drop ${\Pi^\pm}_{ij}^k$ because the terms with the pressures 
$\tilde{\pi}$ and $\tilde{\pi}_s'$ vanish everywhere, as follows from the 
usual arguments of isotropy \cite{KaHo1938}. Finally we obtain
\begin{equation}
\partial_t {\cal Q}_{ij}^\pm = \frac{\partial}{\partial r^k} \left( 
    {{\cal T}^\pm}_{ij}^k - \alpha^2 {{\cal S}^\pm}_{ij}^k \right) .
\label{eq:KH1}
\end{equation}
This is the \lamhda version of eq. (3.8) in \cite{Ho2002c}. In the case 
$\alpha=0$ this equation is also a linear combination of eq. (43), (50), and 
(56) in \cite{Ch1951}. More K\'arm\'an-Howarth equations can be written 
for different combinations of $z^\pm$ and $z_s^\mp$.

Since ${\cal Q}_{ij}^\pm$ and ${{\cal T}^\pm}_{ij}^k$ are symmetric and 
divergence free in their indices $i$ and $j$, ${{\cal S}^\pm}_{ij}^k$ must be 
symmetric and divergence free in $i$ and $j$. But the Els\"asser variables 
$\zvec^\pm$ are combinations of vectors and pseudovectors. Therefore, 
${\cal Q}^\pm$ is a combination of tensors and pseudotensors. We can 
define a tensor as
\begin{equation}
{\cal Q} = {\cal Q}^\pm + {\cal Q}^\mp ,
\end{equation}
and a pseudotensor as ${\cal Q}^\pm - {\cal Q}^\mp$. We will 
continue using only the tensor ${\cal Q}$, the results can also be 
obtained for the pseudotensors using the expressions in \cite{Ch1951}. 
We also define ${\cal T} = {\cal T}^\pm + {\cal T}^\mp$ 
and ${\cal S} = {\cal S}^\pm + {\cal S}^\mp$.

Imposing isotropy and from incompressibility, ${\cal Q}$ can be 
written as \cite{Ch1950}
\begin{equation}
{\cal Q}_{ij} = {\curl} (Q r_l \epsilon_{ijl}) = -(d+1) Q \delta_{ij} + r 
    Q' \left( \frac{r_i r_j}{r} - \delta_{ij} \right) ,
\label{eq:Q}
\end{equation}
where the curl is taken with respect to the third index ($j$), $Q=Q(r,t)$ is 
a scalar function, $\epsilon$ is the Levi-Civita pseudotensor, and $d$ is 
the number of dimensions. Here, $Q' = \partial_r Q$.

In the same way, we can write
\begin{eqnarray}
{\cal T}_{ij}^k &=& {\curl} \left[ T \left( r_i \epsilon_{jkl} r^l + r_j 
    \epsilon_{ikl} r^l \right) \right] \nonumber \\
    &=& \frac{2}{r} T' r_i r_j r^k - (rT' +dT) \left(r_i \delta_j^k + \
    r_j \delta_i^k\right) + 2T \delta_{ij} r^k .
\end{eqnarray}
The tensor ${\cal S}$ takes the same form with scalar function $S(r,t)$. 
Note that ${\cal S}$ is the isotropic sub-$\alpha$-scale stress tensor in 
the LES formulation of {\sl LAMHD$-\alpha$}.
 
Now we compute the divergence of these tensors. In three dimensions
\begin{equation}
\frac{\partial}{\partial r^k} {\cal T}_{ij}^k = {\curl} \left[ (rT' +5T) 
    \epsilon_{ijl} r^l \right] ,
\label{eq:divT}
\end{equation}
and the divergence for ${\cal S}$ takes the same form. Replacing eqs. 
(\ref{eq:Q}) and (\ref{eq:divT}) into eq. (\ref{eq:KH1}) we finally obtain
\begin{theorem}[Karman-Howarth Theorem for \lamhda]\label{KHThm}
$\quad$\\
The exact \lamhda model relation (\ref{eq:KH1}) for homogeneous isotropic
statistics implies the isotropic tensor relation in three dimensions
\begin{equation}
\frac{\partial Q}{\partial t} = \left( r \frac{\partial}{\partial r} + 5 
    \right) \left( T^2 - \alpha^2 S \right) ,
\end{equation}
and in $d$ dimensions the general result is
\begin{equation}
\frac{\partial Q}{\partial t} = \left[ r \frac{\partial}{\partial r} + 
    (d+2) \right] \left( T^2 - \alpha^2 S \right) .
\label{eq:KH2}
\end{equation}
\end{theorem}
This is the generalization of the K\'arm\'an-Howarth equation for \lamhda
(two more equations can be written for different combinations of the 
tensors and pseudotensors), without the dissipation. When $\Bvec=0$ 
this equation is also eq. (3.16) in \cite{Ho2002c}. When $\alpha=0$, this 
is equivalent to the K\'arm\'an-Howarth equation for the Els\"asser variables 
as derived in \cite{PoPu1998}, or a combination of equations (49) and 
(53) in \cite{Ch1951}.

Therefore, all equations in \cite{PoPu1998} follow for $\alpha/r \ll 1$. This 
result confirms that the alpha-model preserves the properties of MHD for 
separations larger than $r \sim \alpha$. For $r>\alpha$, the scaling of 
structure functions and the relation between second and third order 
functions hold.

\begin{corollary}[Kolmogorov Theorem for \lamhda]\label{KThm}
$\quad$\\
Introducing the flux $\partial_t Q = - 2 \epsilon_\alpha /d$ 
with $\epsilon_\alpha = \epsilon_\alpha^+ + \epsilon_\alpha^-$ (the 
energy injection rate for each Els\"asser variable) in eq. (\ref{eq:KH2}) 
and integrating in the inertial range yields
\begin{equation}
-\ \frac{2}{d(d+2)} \epsilon_\alpha = \left( T - \alpha^2 S \right) \ ,
\label{Els-2/d(d+2)Law}
\end{equation}
where $T$ and $S$ are defined in equations (\ref{eqT}) and (\ref{eqS}).
\end{corollary}
Note a multiplicative factor compared with the usual expression from 
Kolmogorov, related to the relation between autocorrelation functions 
and structure functions in isotropic turbulence in $d$ dimensions. For 
$\alpha/r \ll 1$ this equation reduces again to the MHD results. Note 
also that structure and autocorrelation functions 
in \lamhda involve one unsmoothed field and one smoothed field if 
quantities are of second order, and two smoothed fields if quantities 
are of third order. In the following sections, we will use this 
convention and all structure functions for \lamhda will be written as 
they follow from the expressions of the tensors ${\cal Q}$ and ${\cal T}$.

\section{Energy dissipation anomaly}\label{erg-anom-sec}
The K\'arm\'an--Howarth (KH) theorem for fluid turbulence \cite{KaHo1938}
gives the exact analytical relation between the time rate of change of the
second-order two-point velocity correlation function and the gradient of
the third-order two-point velocity correlation function, as derived from
the Navier-Stokes equation for homogeneous, isotropic turbulence. 

Kolmogorov \cite{K41b} used the structure function form of the KH 
equation, to show -- for homogeneous, isotropic and stationary 
turbulence, in the limit $\nu\to0$ of vanishing kinematic viscosity 
-- that the Navier-Stokes equations lead to an exact relationship 
between the third-order structure function and the energy dissipation 
rate, $\overline{\epsilon}$, which scales linearly in the separation, 
$r$.
This is Kolmogorov's 
famous ``four-fifths law.''
 
By assuming self-similarity of scales in the inertial range Kolmogorov
then was able to deduce, in steps that essentially amount to dimensional
analysis, that the second-order structure function must scale as
$r^{2/3}$ and that consequently  the energy spectrum (which is essentially
the Fourier transform of the second-order structure function) must scale
as $k^{-5/3}$.
As noted in \cite{Fr1995}, Kolmogorov's four-fifths law
\begin{quote}
is one of the most important results in fully developed turbulence
because it is both exact and nontrivial. It thus constitutes a kind of
`boundary condition' on theories of turbulence: such theories, to be
acceptable, must either satisfy the four-fifths law, or explicitly violate
the assumptions made in deriving it.
\end{quote}
The two key assumptions in Kolmogorov's derivation of the four-fifths law are
that (1) an inertial range exists in which  the flow is self-similar and (2) the
energy dissipation rate does not change as one takes the limit $\nu\to0$.

The equivalent of the KH equation was derived for the {\sl
LANS$-\alpha$} model in \cite{Ho2002c}. Since the
model relates the Helmholtz smoothed velocity $\mathbf{u}_s$, to the
unsmoothed velocity $\mathbf{u}$, the appropriate correlation functions
that emerge are the second- and third-order two-point correlation
between $\mathbf{u}_s$ and $\mathbf{u}$. Upon following 
Kolmogorov's analysis for isotropic inertial range statistics, the
corollary to the {\sl LANS$-\alpha$} KH-equation is that solutions of the
{\sl LANS$-\alpha$} equations possess two regimes of scaling, depending on
whether the separation distance $r$ is greater, or less than the size
$\alpha$. For $r>\alpha$, the third-order correlation scales like $r$,
thereby recovering Navier-Stokes behavior. In contrast, for $r<\alpha$ the
third-order correlation scales like $r^3$. If self-similarity is then
assumed one finds for $r > \alpha$ that the second-order correlation scales
like $r^{2/3}$, again recovering Navier-Stokes behavior. However, for $r <
\alpha$ the second-order correlation scales like $r^2$. Correspondingly,
the power spectrum for the smoothed velocity ${\bf u}$ has two regimes,
with a transition from
$k^{-5/3}$ for $k\alpha < 1$ to $k^{-3}$ for $k\alpha > 1$. Thus, the
KH-theorem for the {\sl LANS$-\alpha$} model derived in \cite{Ho2002c} is
consistent with the spectral scaling results found for it in 
\cite{FoHoTi2001} by dimensional arguments. We shall apply similar reasoning
to the \lamhda model in two dimensions. 

\subsection*{Differences from MHD turbulence theory for $r<\alpha$}

The second term in the ``$-2/d(d+2)$ Law'' in equation (\ref{Els-2/d(d+2)Law}) 
(the $\alpha^2 \,{\cal S}$ term on the right side) is
reminiscent of the quantity that appears in the corresponding ``$-2$
Law'' for enstrophy cascade in 2D turbulence. The latter expression contains
two powers of enstrophy and one power of velocity. For example, see the
Appendix B of \cite{Ey1996}, where this identity for 2D turbulence is
derived in detail. 

Likewise, the $\alpha^2 \,{\cal S}$ term in (\ref{Els-2/d(d+2)Law}) for \lamhda
in 2D contains two powers of gradients $\nabla\mathbf{z}_s$ and one power of
$\mathbf{z}_s$ without gradient. Consequently, this should be the dominant term
(compared to the first ${\cal T}-$term) for small separations, when
$r<\alpha$. 
If in addition the \lamhda flow is {\it self-similar}, the dominance of the
$\alpha^2 \,{\cal S}$ term in (\ref{Els-2/d(d+2)Law}) when $r<\alpha$  
implies a scaling relation for the second-order structure functions. Following
\cite{K41b} as amplified by
\cite{Fr1995}, let the longitudinal difference
$\delta\mathbf{z}(\mathbf{x},{r})$ satisfy the scaling relation 
$\delta\mathbf{z}(\mathbf{x},\lambda{r})
=\lambda^h\delta\mathbf{z}(\mathbf{x},{r})$ for all $\mathbf{x}$ and for
increments $r$ and $\lambda{r}$ small compared to $\alpha$. By dimensional
analysis, $[{\cal S}(\lambda{r})]=[(\delta\mathbf{z})^3/r^3]=[{\cal S}(r)]$.
Consequently, $3h-3=0$ and $h=1$ for small scales $r<\alpha$
in a self-similar \lamhda flow. This means the {\it second-order}
structure functions follow $r^2$ scaling for $r<\alpha$ in such a
flow. This $r^2$ scaling implies a $k^{-3}$ law for the spectral density of
smoothed kinetic and magnetic energy in that range for the 2D \lamhda model. Thus,
one finds a self-similar $k^{-3}$ ``enstrophy-like" cascade, in agreement with
similar considerations of \cite{FoHoTi2001} for the kinetic energy spectral
density in  the 3D LANS$-\alpha$ model. 

\paragraph{Implications of the $k^{-5/3}\to k^{-3}$ spectral scaling
transition for the \lamhda dissipation anomaly.} 
The second term in the ``$-2/d(d+2)$ Law'' in equation (\ref{Els-2/d(d+2)Law})
modifies Kolmogorov's four-fifths
law at small separations ($r<\alpha$), provided one may assume constancy of total
\lamhda energy dissipation as $\nu\to0$. This is the ``energy
dissipation anomaly'' for the \lamhda model. A technical argument
using embedding theorems for Besov spaces first introduced in
\cite{CoETi1994Ey2002} implies that constancy of total
energy dissipation may hold as $\nu\to0$ for a turbulent fluid in two
dimensions, provided its $L^2$ power spectrum is not steeper  than $k^{-4}.$ The
$k^{-3}$ spectrum for $k\alpha > 1$ is not too steep;  so the roll-off
$k^{-5/3}\to k^{-3}$ in the \lamhda power spectrum is consistent with the
necessary condition for possessing such  an energy dissipation anomaly. Hence,
the $k^{-3}$ behavior in the  $L^2$ power spectrum of the \lamhda model for
$k\alpha>1$ and the corresponding modification for separations $r<\alpha$ of
Kolmogorov's four-fifths law derived in \cite{Ho2002c} are both consistent with
the  assumption of constant dissipation of {\it total} kinetic energy as
the Reynolds number tends to infinity. 

\paragraph{Implications of the spectral scaling transition 
for IK scaling instead of Kolmogorov scaling.}

While in hydrodynamic turbulence the spectral transfer is believed to 
be a local process, in MHD turbulence Iroshnikov and Kraichnan proposed 
that the spectral transfer is governed by nonlocal Alfv\'en wave 
interactions (see \cite{Alex} for a study of non-locality of transfer in MHD).
If a large scale magnetic field is present, then this 
field acts as a guide field to the fluctuations, turning them into 
Alfv\'en waves. Kraichnan proposed that Alfvenic propagation limits 
the nonlinear interaction responsible for the transfer of energy to 
smaller eddies in the absence of magnetic fields. As a result, the IK 
spectrum is shallower than K41. In {\lamhda}, the $k^{-3}$ spectrum 
at scales smaller than $\alpha$ turns into a shallower $k^{-5/2}$ 
when the IK hypotheses are used. Note that the weakening of local 
interactions due to Alfv\'en waves holds in \lamhda at the smoothed 
scales, and the energy dissipation anomaly should also be captured. 
This is also in agreement with equations (\ref{eq:Elsasser1}) and 
(\ref{eq:Elsasser2}), which show that Alfv\'en waves are also exact 
nonlinear solutions of the \lamhda equations.

\section{Numerical results for intermittency and
scaling anomaly}\label{scale-anom-sec}

In this section, we compare intermittency in \lamhda to that of 
direct numerical simulations (DNS) of the MHD equations, regarded 
as true at a given Reynolds number. Intermittency is associated both with the presence of strong 
localized structures and with the existence of strong non-Gaussian 
wings in the probability distribution functions. We have previously 
investigated the latter in \cite{lamhd2d} and we concentrate here on the 
strong localized structures giving rise to deviations from universality, 
as can be studied by examining high order statistical moments, such as 
the structure functions. 

We define the longitudinal structure function of a field {\bf f} as 
$\mathfrak{S}_p^f(l) \equiv \langle|\delta f_L|^p\rangle$ where 
$\delta f_L = ({\bf f}({\bf x} + {\bf l}) - {\bf f}({\bf x})) \cdot
{\bf l}/l$ is the longitudinal increment of ${\bf f}$. In the inertial 
range between the large energy-containing scales and the small 
dissipative scales, the structure functions are assumed to vary in a
self-similar manner, $\mathfrak{S}_p^f(l) \sim l^{\zeta_p^f}$. As previously 
mentioned, in isotropic and homogeneous turbulence the structure functions 
can be related to the correlation functions discussed in Section 
\ref{KH-sec}. K41 phenomenology predicts $\zeta_p^v = p/3$, while IK 
gives $\zeta_p^\pm = p/4$.  The anomalous departure of the exponents 
$\zeta_p$ from these linear scaling laws is a measure of 
intermittency-induced deviations from universality.

To numerically solve the MHD and \lamhda equations we will use a parallel 
pseudospectral code as described in \cite{lamhd2d}. In two dimensions, the 
velocity and magnetic field are expressed as the curl of a scalar stream 
function $\Psi$ and a one component vector potential $A_z$, respectively:
   \begin{equation}
     \begin{array}{lr}
     \vvec = {\bf \nabla} \times (\Psi {\bf \hat{z}}),  &
     \uvec = {\bf \nabla} \times (\Psi_s {\bf \hat{z}})
     \end{array}
   \end{equation}
   \begin{equation}
     \begin{array}{lr}
     {\bf B} = {\bf \nabla} \times (A_z {\bf \hat{z}}),  &
{\bf B}_s = {\bf \nabla} \times (A_{s_z} {\bf \hat{z}})
     \end{array}
     \label{EPSILON}
   \end{equation}
where $\Psi = (1-\alpha^2 \nabla^2) \Psi_s$, and 
$A_z = (1-\alpha_M^2 \nabla^2) A_{s_z}$. In terms of these quantities, the 2D MHD
equations may be expressed as
   \begin{equation}
     \partial_t \nabla^2 \Psi = [\Psi,\nabla^2\Psi] - [A_z,\nabla^2A_z] +
     \nu \nabla^2\nabla^2 \Psi
   \end{equation}
   \begin{equation}
     \partial_t A_{z} = [\Psi,A_{z}] + \eta \nabla^2 A_z,
   \label{2DMHD}
   \end{equation}
where
   \begin{equation}
     [F,G]= \partial_xF\partial_yG - \partial_xG\partial_yF 
   \end{equation}
is the standard Poisson bracket. The \lamhda  equations
(\ref{eq:lamhd1}-\ref{eq:lamhd3}) modify this two dimensional structure by
introducing smoothed variables as 
   \begin{equation}
     \partial_t \nabla^2 \Psi = [\Psi_s,\nabla^2\Psi] - [A_{s_z},\nabla^2A_z] +
     \nu \nabla^2\nabla^2 \Psi
   \end{equation}
   \begin{equation}
     \partial_t A_{s_z} = [\Psi_s,A_{s_z}] + \eta \nabla^2 A_z
\,.
   \label{2DLAMHD}
   \end{equation}
In the following subsections we test the \lamhda model against MHD results
(for which $\alpha = \alpha_M = 0$) for freely decaying turbulence with the 
same initial conditions, dissipation and time-stepping, and also for 
forced turbulence where we have averaged statistics over 189 turnover 
times taken from 9 experiments with distinct seeds for the random 
forcing, resulting in a data set of $\sim 2\cdot 10^8$ points.

\subsection{Forced simulations}

In this subsection we consider forced turbulence with 
$\eta = \nu = 1.6 \times 10^{-4}$. Four sets of simulations were 
carried out, one set of MHD fully-resolved simulations with $1024^2$ 
grid points, and three sets of \lamhda simulations, with $512^2$ grid 
points and $\alpha=\alpha_M=6/512$, with $256^2$ grid points and 
$\alpha=\alpha_M=6/256$, and with $256^2$ grid points and 
$\alpha=\alpha_M=6/128$. Note that the $256^2$ \lamhda simulation 
with $\alpha=\alpha_M=6/128$ could be carried out with a $128^2$ spatial 
resolution (see e.g. criteria for the selection of the value of $\alpha$ 
and the linear resolution in \cite{lamhd2d}). The reason to keep the 
resolution fixed at $256^2$ is to preserve the amount of spatial 
statistics, crucial to measure high order exponents as will be shown 
later.

Both the momentum and the vector potential equations were forced. The 
expressions of the external forces were loaded in the Fourier ring 
between $k=1$ and $k=2$, and the phases were changed randomly with a 
correlation time $\Delta t = 5 \times 10^{-2}$. Averaged over space, 
the amplitudes of the external forces were held constant to $F_M = 0.2$ 
in the vector potential equation, and $F_K = 0.45$ in the momentum 
equation. The systems were evolved in time until reaching a turbulent 
steady state, and then the simulations were extended for 21 turnover 
times. Over this time span, 21 snapshots of the fields from each run 
were used to compute the longitudinal increments.  As previously 
mentioned, each set of simulations (DNS, and \lamhda with different 
spatial resolutions) comprises nine runs with the same viscosity and 
diffusivity but different series of random phases in the external 
forcing, to have enough statistics to determine the scaling exponents 
up to eighth order with small error bars. The total number of points 
was $\sim2\cdot10^8$ for the DNS and $\sim1.2\cdot10^7$ points for 
the $256^2$ \lamhda simulations.

During these intervals, the integral Reynolds number based on the {\it r.m.s.}
velocity fluctuates around 2200.  The normalized correlation coefficient between the velocity and the magnetic field is 20\% 
with a standard deviation of 16\% within the set of nine runs, and 
its unsigned counterpart 
$2\langle|\vvec\cdot\Bvec|\rangle / \langle|\vvec|^2+|\Bvec|^2\rangle$,
is $\sim29\%\pm12\%$.  The ratio of the integral scale to the Taylor scale
computed on the {\it r.m.s.} fields fluctuates around 10 for all the simulations. 
The ratio of magnetic to kinetic energies is $\sim2$ for all runs. Finally, 
the Kolmogorov dissipation wavenumbers 
$k_\nu = (\left<\omega^2\right>/\nu^2)^{1/4}$ and 
$k_\eta = (\left<j^2\right>/\eta^2)^{1/4}$ fluctuate around $330$, values
substantially larger than the largest resolved wavenumbers $k_\alpha \sim 1/\alpha$ 
in all \lamhda simulations, by virtue of the model.

\begin{figure}
  \begin{center}	
  \includegraphics[width=10cm]{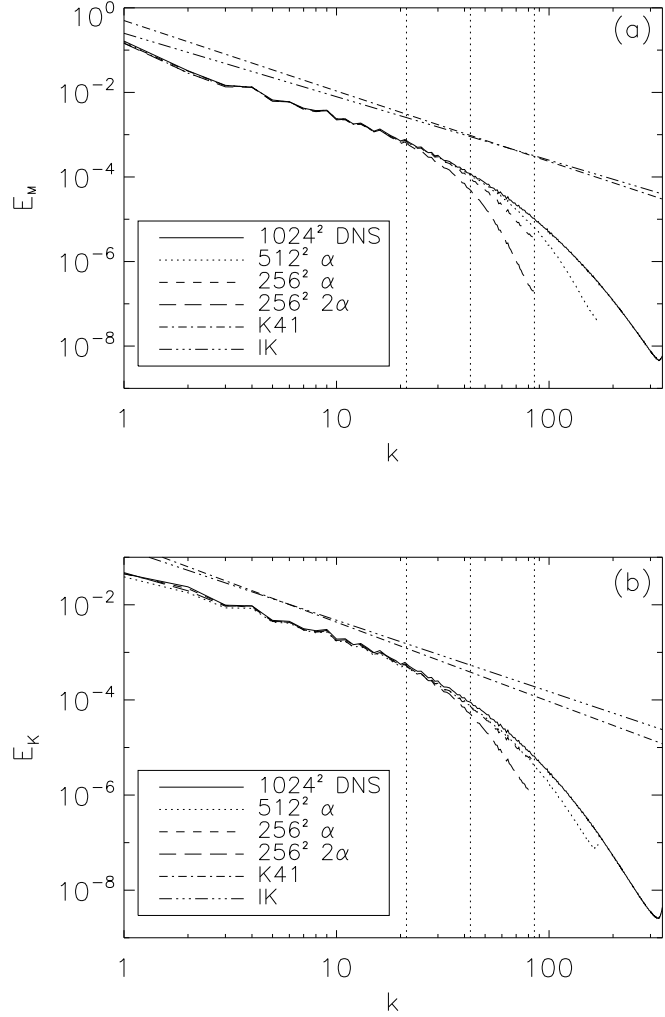}
    \end{center}
  \caption {Spectra averaged over 189 turnover times. $1024^2$ MHD is 
  the solid line, $512^2~$ \lamhda is the dotted line, $256^2$ \lamhda 
  is the dashed line, $256^2$ \lamhda with $\alpha = \alpha_M = 6/128$ 
  is the long-dashed line (hereafter indicated in figures as 
  '$256^2$ $2\alpha$'), $k^{-5/3}$ (K41) is the dash-dotted 
  line, and $k^{-3/2}$ (IK) is the dash-triple-dotted line; the 
  K41 and IK slopes are shown for reference. The vertical lines indicate 
  the wavenumbers corresponding to the lengths $\alpha$ for all 
  \lamhda simulations. Panel {\bf (a)} is magnetic energy, $E_M$, 
  versus wavenumber $k$, and panel {\bf (b)} is kinetic energy, $E_K$ 
 {\it vs.} $k$.}
  \label{FIG7}
\end{figure}

Average omni-directional spectra for magnetic and kinetic energies over 
these 189 turnover times are shown in Figure \ref{FIG7}. 
All spectra display an inertial range, and the \lamhda simulations are 
able to capture the spectral behavior up to $k \approx 1/\alpha$.
\neu{}{For $k > 1/\alpha$, theoretical arguments suggest a $k^{-3}$
spectrum for the alpha model.  To observe this spectrum, however, large 
values of $\alpha$ would be required. This is inconsistent with the use 
of the alpha model as a subgrid turbulent model, and lies beyond the 
interest of the present work.}

\neu{}{K41 theory predicts
\begin{equation}
\mathfrak{S}_p^u(l) \sim l^{\zeta_p^u},
\label{EQ:revision1}
\end{equation}
which follows from the assumption that the statistical properties of the field
are self-similar in the inertial range, which will be identified here as the 
scales for which the relation
\begin{equation}
{\zeta_3^u} = 1
\label{EQ:revision2}
\end{equation}
holds.  The existence of scaling (\ref{EQ:revision1}) has been extensively verified for the
hydrodynamic case.  Starting from the
assumption of self-similarity in the inertial range, we can then postulate the validity of (\ref{EQ:revision1}) at
arbitrarily high order, $p$.  From an experimental standpoint, the amount of
data used in calculating $\mathfrak{S}_p^u(l)$ determines the highest order
for which we can observe a scaling in the inertial range.  Above this order,
the assumption of self-similarity allows us to fit a scaling law to our data
in any event.  But in the absence of sufficient statistics, such a fit can be 
poor and the error bars rather large (see, {\sl e.g.}, Figure \ref{FIG21} and
Figure \ref{FIG_COMPARE} {\bf (a)} to be discussed shortly).

The Extended Self-Similarity (ESS) hypothesis \cite{BCB+93}, proposes the
scaling
\begin{equation}
\mathfrak{S}_p^u(l) \sim \left[  \mathfrak{S}_3^u(l) \right]^{\xi_p^u},
\label{EQ:revision3}
\end{equation}
which is found to apply to a much wider scaling range than the inertial range.
Here, the scaling range is determined by the observed scaling for low orders and
a chief benefit is increased statistics to compute more accurate
exponents at higher order.
For the case of MHD, Ref. \cite{PoPu1998} proposes to replace
$\mathfrak{S}_3^u(l)$ with the third-order, mixed structure 
functions, $L^\pm(l)$.  As in the ESS hypothesis, these scales $L^\pm(l)$}
may provide better independent variables (as opposed to length {$l$}) 
against which to determine the scaling exponents
for MHD. From the KH theorem for an incompressible, non-helical 
MHD flow, they find
\begin{equation}
\langle\delta z_L^\mp(\vec{l})|\delta z^\pm(\vec{l})|^2\rangle = 
   - \frac{4}{d}\varepsilon^\pm l,
\end{equation}
where $|\delta z^\pm|^2 = ({\delta z_L^\pm})^2+({\delta z_T^\pm})^2$, 
$\delta z_T^\pm$ are the transverse increments, $d$ is the space dimension,
$\varepsilon^+$ and $\varepsilon^-$ are the energy dissipation rates for 
$\frac{1}{2}({z^+})^2$ and $\frac{1}{2}({z^-})^2$ respectively, and 
angle brackets indicate as usual spatial averages \cite{PoPu1998}.

These results for the third-order structure functions are exact
and can be used to compute more accurate anomalous scaling exponents 
of structure functions of higher order.  Due to cancellation problems (linked
with having limited statistics), absolute values are often employed; we also find linear scaling in this case, {\it viz.}:
\begin{equation}
  L^\pm(l) \equiv \langle|\delta z_L^\mp||\delta z^\pm|^2\rangle\propto l \ .
  \label{THIRDORD}
\end{equation}
 As follows from the expressions given
in Sec. \ref{KH-sec} and the invariants found for both MHD and 
{\sl LAMHD$-\alpha$} \cite{lamhd2d,Ho2002}, when making 
comparisons between DNS and model runs, we substitute the $H^1_\alpha$ norm, 
$\langle||u||_{\alpha}^2\rangle=\langle|\vec{u}\cdot\vec{u}_s|\rangle$
\cite{HoMaRa1998b,HoMaRa1998a}, for the regular $L^2$ norm,
$\langle|u|^2\rangle=\langle|\vec{u}\cdot\vec{u}|\rangle$,
whenever we consider quantities for the \lamhda model.
\neu{}{The Karman-Howarth theorem for {\sl LAMHD$-\alpha$} is essential to this
study of intermittency in that it allows us to define the structure functions
for {\sl LAMHD$-\alpha$}; it also identifies the flux relation that scales linearly with $l$
for application in MHD of the ESS hypothesis.}
 Accordingly 
we determine the relative scaling exponents, $\xi_p^f$, by using Eq. 
(\ref{THIRDORD}) for the third-order, mixed structure function, 
$L^+(l) = \langle|\delta z_L^-||\delta z^+|^2\rangle$ for MHD and
$L_s^+(l)=\langle|\delta z_{s_L}^-|||\delta z^+||_{\alpha}^2\rangle$ for
LAMHD-$\alpha$,
\begin{equation}
  \mathfrak{S}_p^f(l) \sim [L_{(s)}^+(l)]^{\xi_p^f}.
  \label{RESCALING}
\end{equation}

\begin {figure}
    \includegraphics[width=11cm]{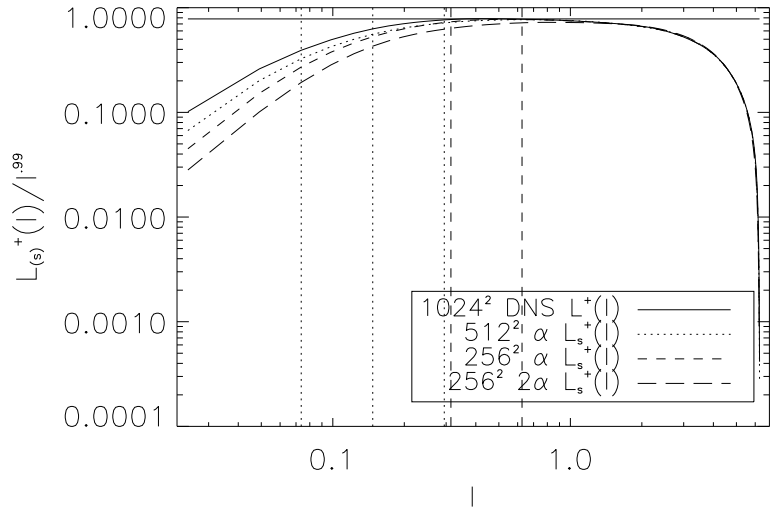}
  \caption {Third-order, mixed structure function, $L_{(s)}^+(l)/l^{0.99}$, versus l, 
  for forced runs of turbulence averaged over 189 turnover times. Results are scaled 
  by the mean value of $L_{(s)}^+(l)$ for easier comparison. Labels 
  are as in Fig. \ref{FIG7}. The best fit to the MHD data, $L^+\sim l^{0.99}$, is 
  indicated by the solid straight line.  The inertial range where this fit is 
  made is indicated by dashed vertical lines and dotted vertical lines 
  indicate the lengths $\alpha$ for the $512^2$ and $256^2$ simulations.}
  \label{FIG8}
\end{figure}

Figure \ref{FIG8} shows the third-order mixed structure function 
$L_{(s)}^+(l)/l$ as a function of $l$. We find, contrary to what is reported in \cite{biskamp},
that the relation (\ref{THIRDORD}) has a identifiable range of validity, as can be seen in the figure 
by comparison with the solid straight line denoting the computed slope 
$L^+(l) \sim l^{0.99}$.  \neu{}{This range of validity is identified as the
inertial range, $2\pi/20 \leq l \leq 2\pi/10$, 
indicated by dashed vertical lines.} The \lamhda runs display the same scaling 
as the MHD simulation, and \neu{}{departures are pronounced only for scales
approaching and
smaller than $\alpha$ (for the $256^2$ runs, $\alpha\approx0.15$, $0.29$ 
and for the
$512^2$ run, $\alpha\approx0.07$ as indicated by dotted vertical lines).}
 Note that the 
results have been scaled by the mean value of $L_{(s)}^+(l)$.  As the 
average energies of the runs are disparate, this improves the ease of 
comparison. The same behavior is observed for $L_{(s)}^-(l)$.

\begin{figure}
  \begin{center}
    \includegraphics[width=11cm]{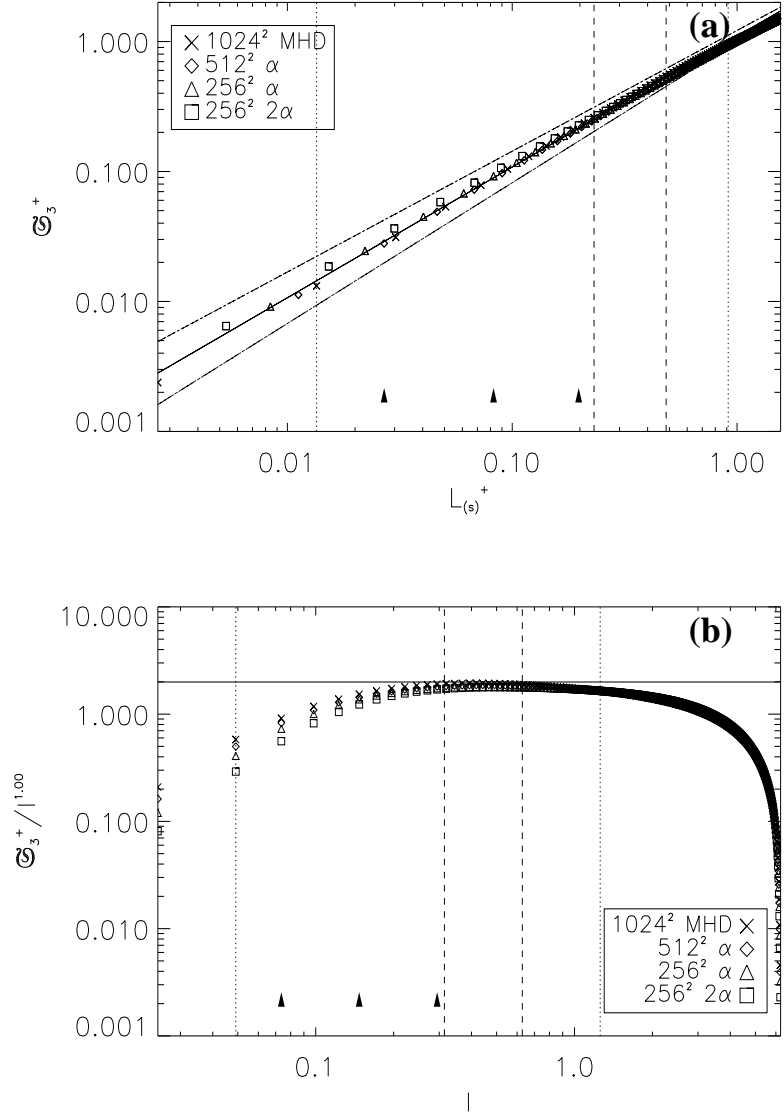}
  \end{center}
  \caption {Third-order structure functions for $\vec{z}^+$: Panel {\bf(a)} 
  $\mathfrak S_3^+$ versus $L_{(s)}^+$ and panel {\bf(b)}
  $\mathfrak S_3^+/{l}^{\zeta_3^+}$ versus $l$, computed over 189 turnover times.  Labels 
  are as in Fig. \ref{FIG7}. The solid line corresponds to the best fit to the DNS data,
  $\mathfrak S_3^+ = (L^+)^{1.01}$ (the  dash-dotted lines represent the
  3$\sigma$ error).  The ESS hypothesis range where this fit is 
  made is indicated by dotted vertical lines and dashed vertical lines 
  indicate the inertial range. Arrows indicate the several lengths 
  $\alpha$ used in the simulations.}
  \label{FIG9}
\end{figure}

\begin{figure}
  \begin{center}
    \includegraphics[width=11cm]{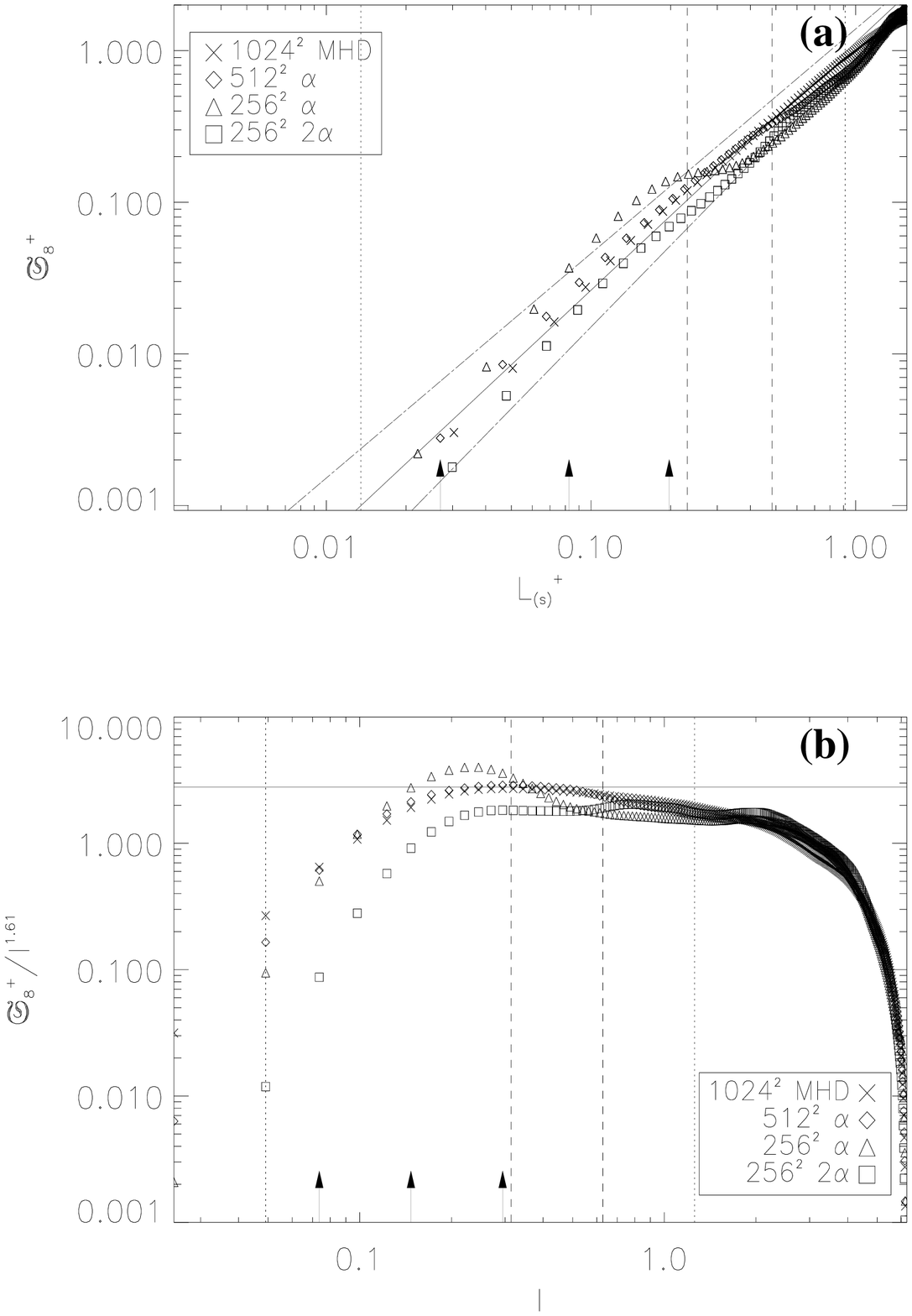}
  \end{center}
  \caption {Eighth-order structure functions for $\vec{z}^+$: Panel {\bf(a)} 
  $\mathfrak S_8^+$ versus $L_{(s)}^+$ and panel {\bf(b)}
  $\mathfrak S_8^+/{l}^{\zeta_8^+}$ versus $l$, computed over 189 turnover times. Labels 
  are as in Fig. \ref{FIG9}. The solid line corresponds to the best fit to the
  DNS data,
  $\mathfrak S_8^+ = (L^+)^{1.63}$.  The ESS hypothesis range where this fit is 
  made is indicated by dotted vertical lines and dashed vertical lines 
  indicate the inertial range. Arrows indicate lengths $\alpha$.}
  \label{FIG21}
\end{figure}

The scaling of the third-order structure function
$\mathfrak S_3^+$ 
versus $L_{(s)}^+$ for the Els\"asser variable $\vec{z}^+$ is shown 
in Figure \ref{FIG9} \neu{}{{\bf(a)} as well as a compensated plot versus $l$ in
 Figure \ref{FIG9} {\bf(b)}.  We see very little contamination at scales larger
than $\alpha$.  This contrasts with hyperviscosity which is known to cause an
enhanced bottleneck in Navier-Stokes turbulence, which corrupts the scaling of the larger scales for
structure functions of order two and higher \cite{HB04}.  The MHD case has not been studied in this context, and the presence of an inverse cascade of magnetic helicity might also exacerbate the problem.}

 The \lamhda simulations show similar scaling to what is found in DNS of MHD. 
A solid line indicates the best fit to the DNS data, $\xi_3^+ = 1.01 \pm 0.08$ 
using the ESS hypothesis \neu{}{(the ESS scaling range is} indicated by dotted vertical lines); note that
all errors presented and shown in the figures correspond to $3\sigma$ 
where $\sigma$ is the standard deviation.
The much smaller inertial range 
is indicated by dashed vertical lines and arrows indicate lengths $\alpha$.
This is the main benefit of using the ESS hypothesis. While in the $1024^2$ 
simulation there is enough statistics to measure the scaling exponents 
$\xi_p$ in the inertial range, when we use the \lamhda equations to 
reduce the computational cost, the amount of spatial statistics is 
drastically reduced ({\it e.g.} by a factor of 16 in the $256^2$ runs). The 
ESS hypothesis allows us to extend the range where the $\xi_p$ 
exponents are computed, giving a better estimation and smaller error 
bars. 
As an example, in 
Fig. \ref{FIG21} {\bf(a)} we show the scaling of the eighth-order structure 
function $\mathfrak S_8^+$ versus $L_{(s)}^+$ \neu{}{as well as a compensated plot versus $l$ in
 Figure \ref{FIG21} {\bf(b)}.} The ranges 
corresponding to the inertial range and ESS are also indicated.
\neu{}{For the $256^2$ runs, we cannot observe a scaling at this order.
The error of a scaling computed from the assumption of self-similarity is
excessively large (see  Figure \ref{FIG_COMPARE} {\bf (a)}).
  From the ESS hypothesis a better estimation of the (postulated)
scaling at order eight can be made (see  Figure \ref{FIG_COMPARE} {\bf (b)}).}

Figure \ref{FIG_COMPARE} compares the scaling exponents, $\xi_p^+$, for 
the DNS runs and the three sets of \lamhda runs.  Figure \ref{FIG_COMPARE} 
{\bf (a)} is for exponents computed only over the inertial range. Notice that
the \lamhda runs capture the scaling for the low-order moments ($p \leq 4$). 
For higher-order moments (beginning at $p=5$), the drop in the scaling 
exponents for the $256^2$ results (with $\alpha=\alpha_M=6/256$) and the 
large error bars are indicative of insufficient statistics.
\neu{}{The advantages of ESS are clearly seen by comparison with}
Figure \ref{FIG_COMPARE} {\bf (b)}, which shows the 
scaling exponents for all sets of runs employing the ESS hypothesis. 
In both figures, the She-L\'ev\^eque (SL) formula \cite{She94} modified 
for the MHD case \cite{PoPu1995} is shown as a reference,
\begin{equation}
\frac{\xi_p}{\xi_3} = \frac{p}{6} + 1 -\left(\frac{1}{2}\right)^{p/3} .
\label{eq:sheleveque}
\end{equation}
From these results, we conclude that \lamhda captures the intermittency 
of the DNS runs up to and including the eighth-order moment (to within 
the errors of our statistics).
\neu{}{The size of the error bars make it difficult to draw further conclusions
from the data.}

\begin{figure}
  \begin{center}
  \leavevmode
   \includegraphics[width=10cm]{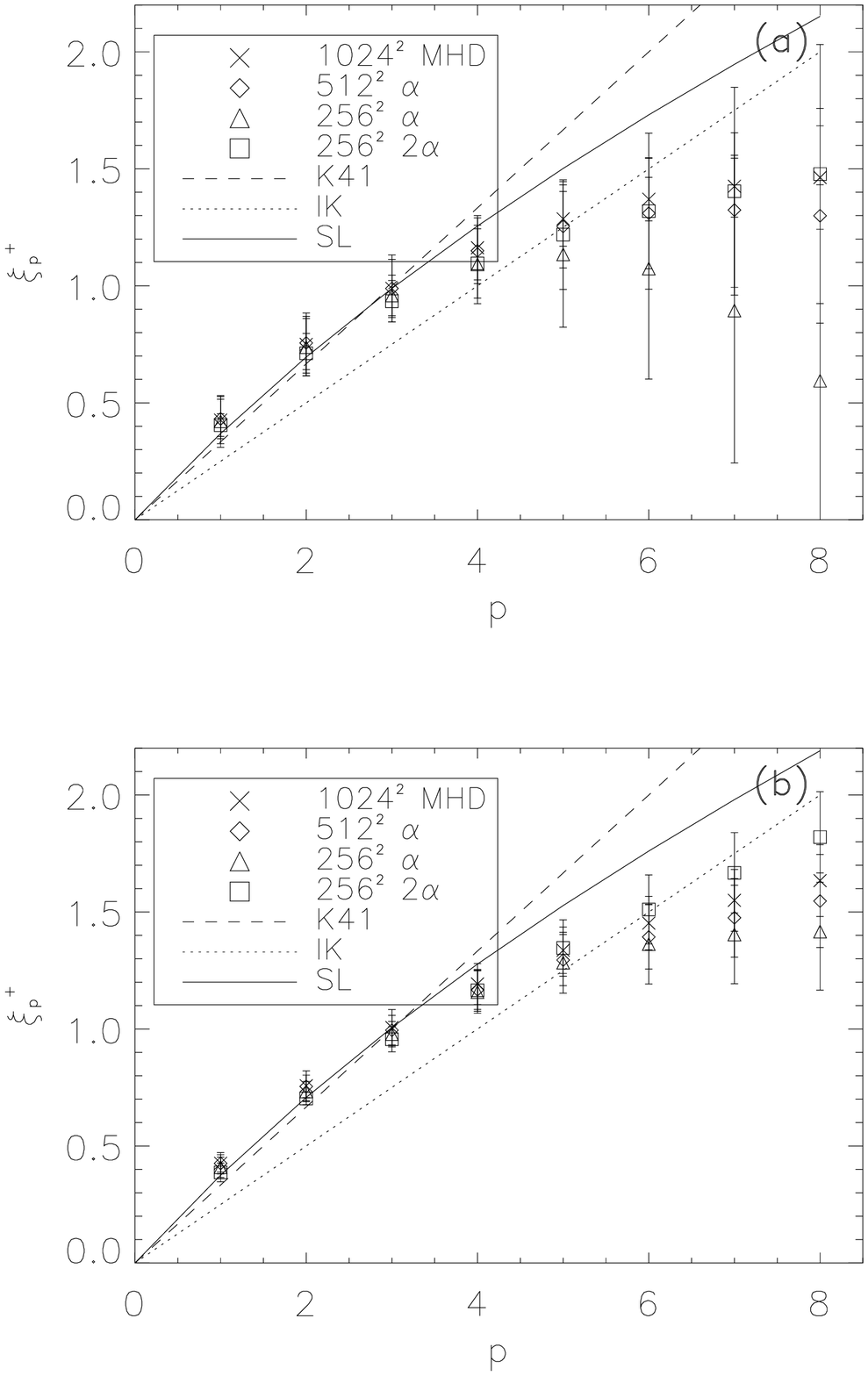}
  \caption {Structure function scaling exponent: $\xi_p^+$ versus $p$, 
  computed over 189 turnover times. $1024^2$ MHD 
  are the pluses, $512^2$ \lamhda are the diamonds, $256^2$ \lamhda 
  ($\alpha=6/256$) are triangles, and $256^2$ \lamhda ($\alpha=6/128$) are 
  the squares. The dashed line indicates K41 scaling, 
  dotted line indicates IK scaling, and the solid line is the prediction 
  using the modified She-L\'ev\^eque formula (see text). Panel {\bf (a)} is computed over the 
  inertial range. Panel {\bf (b)} is computed utilizing the ESS hypothesis.}
  \label{FIG_COMPARE}
  \end{center}
\end{figure}

\begin{table}[htpb]
\caption{Relative scaling exponents (together with 3$\sigma$ errors in computing the
	slope in parenthesis) computed from the 9 MHD runs
	over the inertial range, $\xi$, and utilizing the ESS hypothesis,
	$\xi_{ESS}$.}
\begin{center}
\begin{tabular}{||l|c|c|c|c|c||} \hline
	$p$ & $\xi^+$  & $\xi_{ESS}^+$  & $\xi_{ESS}^-$  & $\xi_{ESS}^u$  & 
$\xi_{ESS}^B$ \\  \hline \hline
	1 & .43(10) & .43(05) & .42(04) & .37(06) & .43(04) \\ \hline
	2 & .75(13) & .76(06) & .76(04) & .67(08) & .76(04) \\ \hline
	3 & .99(14) & 1.01(08) & 1.01(03) & .92(08) & 1.01(04) \\ \hline
	4 & 1.16(14) & 1.19(09) & 1.20(04) & 1.10(08) & 1.16(02) \\ \hline
	5 & 1.29(11) & 1.34(10) & 1.34(07) & 1.25(08) & 1.27(04) \\ \hline
	6 & 1.37(09) & 1.45(11) & 1.45(09) & 1.36(05) & 1.33(06) \\ \hline
	7 & 1.43(13) & 1.55(13) & 1.53(12) & 1.44(11) & 1.38(07) \\ \hline
	8 & 1.46(22) & 1.63(15) & 1.60(14) & 1.50(13) & 1.43(08) \\ \hline
	\end{tabular}
\end{center}
\label{TABLE1}
\end{table}

The anomalous scaling results for the DNS runs are shown in 
Table \ref{TABLE1}. Though our goal here is to test \lamhda against DNS, 
we remark briefly on the correspondence between our scaling exponents 
and other studies. In opposition to the findings of
\cite{PPC98}, we find $\xi_3^\pm\sim1$ but $\xi_4^\pm>1$.  
We note, however, that the forcing in \cite{PPC98} was tailored to maintain
at a constant level all Fourier modes with $k=1$ while our forcing is random
with a constant amplitude between $k=1$ and $k=2$. As can be seen from 
Table \ref{TABLE1}, our results for $\xi_p^+$ are in good agreement with 
\cite{biskamp}, for decaying turbulence, and with \cite{GoPoPo1999}, for 
forced turbulence. As \cite{biskamp} suggests, the scaling exponents, as 
inertial range properties, may depend on the character of the driving due 
to non-local processes in the cascade dynamics connected with the Alfv\'en 
effect (see also \cite{Alex}); and as Ref. \cite{PPC98} points out, such 
an analysis can be sensitive to several parameters such as the ratio of kinetic to magnetic energy or the
amount of correlation between the velocity and the magnetic fields.

\begin{figure}
  \begin{center}
  \leavevmode
    \includegraphics[width=8cm]{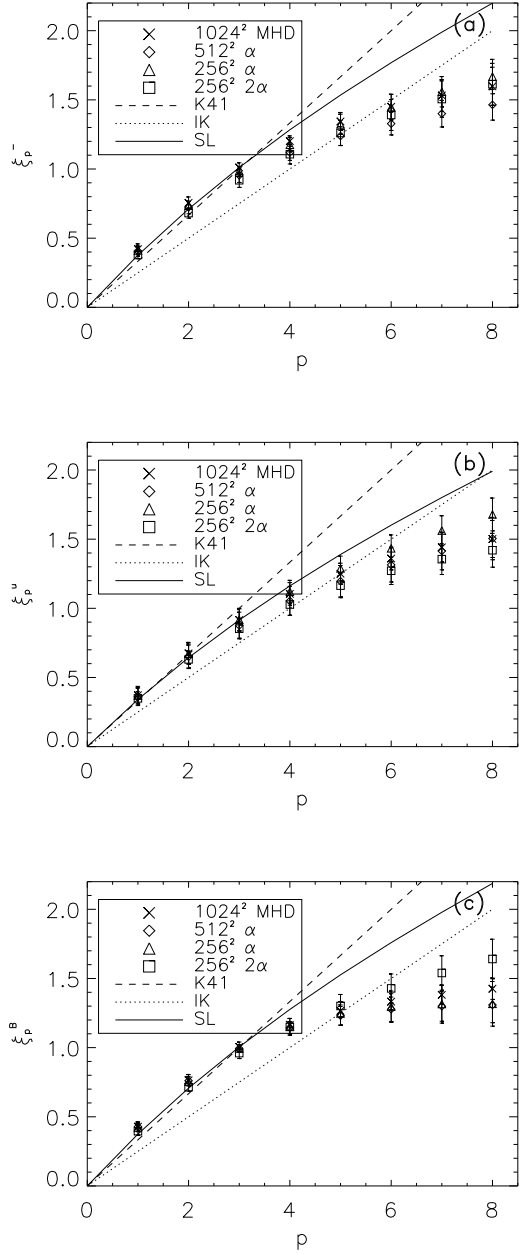}
  \caption {Structure-function scaling exponent $\xi_p^+$ versus $p$ 
  for $\zvec^-$ (panel {\bf(a)}), $\vvec$ (panel {\bf(b)}), and $\Bvec$
  (panel {\bf(c)}). Labels are as in Figure \ref{FIG_COMPARE}.}
  \label{FIG_ESS}
  \end{center}
\end{figure}

Figure \ref{FIG_ESS} shows the scaling exponents for the velocity and 
magnetic fields, as well as for the other Els\"asser variable $\zvec^-$.
The anomalous scaling is again matched by \lamhda up to and including eighth-order. 
The results from the MHD and \lamhda simulations are also in good agreement 
with Ref. \cite{GoPoPo1999}. Note that the magnetic field is more 
intermittent than the velocity field (in the sense that the scaling 
exponents deviate more from a straight line), as previously found in 
numerical simulations \cite{PPC98}; it may be related to the fact that in MHD, nonlinear interactions are more non-local (in Fourier space) than for fluids \cite{Alex}.
This well known feature of MHD 
turbulence is also properly captured by the \lamhda equations.
\neu{}{The average (over all fields) of the $3\sigma$ errors of the eighth
order scaling exponent
is 0.13 for the DNS on a $1024^2$ grid; it is 0.15 and 0.16 for
{\sl LAMHD$-\alpha$} on a $512^2$ and $256^2$ grid, respectively.  To further
test the convergence of our statistics, we reduced the amount
of data used to compute the scaling exponents for the $512^2$ runs by a factor
of 4 (which gives the same amount of statistics than the $256^3$ LAMHD runs) 
and determined an average  $3\sigma$ error of 0.17.  While these results
confirm our convergence as the amount of statistics is increased, they also
highlight the rather low decrease in error with increased computational effort.
Accordingly, the computational burden for more accurate determination of high-order
statistics prohibits further analysis of the data.}

\subsection{Decaying simulations}

In this subsection we briefly discuss simulations of free decaying 
turbulence using both the MHD and the \lamhda equations. The results are 
similar to the ones presented in the previous subsection for forced 
turbulence. However, since no turbulent steady state can be defined in 
freely decaying runs, the amount of statistics is reduced as only 
a few snapshots of the velocity and magnetic field during the peak of 
mechanic and magnetic dissipation can be used to compute structure 
functions. To partially overcome this problem, we will discuss simulations 
with higher spatial resolution than the ones presented in the forced case.

A fully resolved $2048^2$ MHD run was made using $\nu = \eta = 10^{-4}$, 
as well as a $1024^2$ \lamhda run with $\alpha = \alpha_M = 6/1024$ 
and  a $512^2$ \lamhda run with $\alpha = \alpha_M = 6/512$. The initial 
velocity and magnetic fields were loaded with random phases into the rings 
from $k=1$ to $k=3$ in Fourier space. The initial {\it r.m.s.} values of 
$\vvec$ and $\Bvec$ are equal to unity. No external forces are applied and the 
system decays freely as a result of the dissipation. Under these conditions, 
the \lamhda equations have been shown to reproduce the time evolution of the 
magnetic and kinetic energy, as well as the evolution of the spectra and 
other statistical quantities \cite{lamhd2d}. 

The magnetic and kinetic energy spectra between $t=3$ and $t=6$ in units of the
eddy turnover time (the time for which a quasi-steady state is 
observed) display an inertial range with an extent of approximately one 
decade in Fourier space, from $k\approx3$ up to $k\approx30$. As a result, 
one snapshot of the fields in this range in time was used to compute the 
longitudinal structure functions. The Kolmogorov kinetic and magnetic 
dissipation wavenumbers $k_\nu$ and $k_\eta$ peaked at a value of 470, 
larger than the filtering wavenumber $k_\alpha \sim 1/\alpha$ in 
all the \lamhda simulations. We note that for large wavelength component 
behavior up to $k \sim k_\alpha$, both \lamhda simulations accurately 
reproduced the omni-directional spectra for the magnetic and 
kinetic energies as was shown previously for the forced runs.

\begin{figure}
  \begin{center}
    \includegraphics[width=11cm]{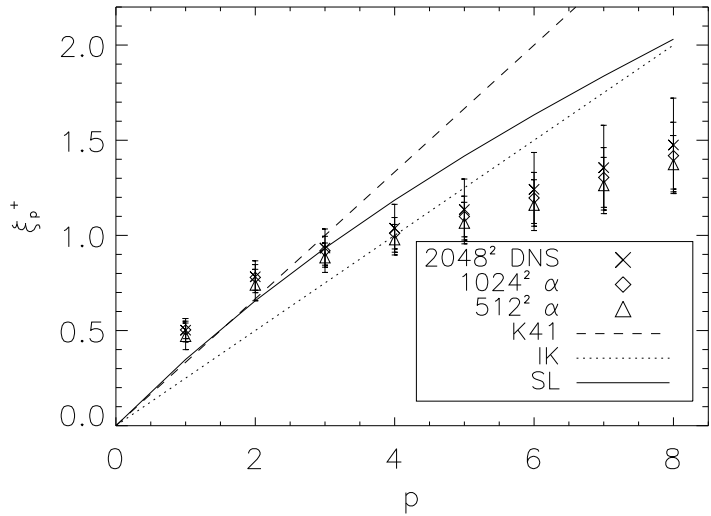}
  \end{center}
  \caption {Scaling exponent $\xi_p^+$ as a function of the order $p$, 
  for the $\zvec^+$ Els\"asser variable in simulations of freely-decaying 
  turbulence. The MHD simulation, the $1024^2$ \lamhda run, and the 
  $512^2$ \lamhda run are denoted by crosses, diamonds, and triangles 
  respectively. The K41 scaling is the dashed line, the IK prediction is 
  the dotted line, and the solid line corresponds to eq. 
  (\ref{eq:sheleveque}).}
  \label{FIG5Z-}
\end{figure}

Both \lamhda runs preserve the scaling of the longitudinal structure 
function exponents observed in the MHD simulation. As an example, 
Fig. \ref{FIG5Z-} shows the $\xi_p$ exponents for the $\zvec^+$ 
Els\"asser variable for the MHD and \lamhda simulations using the 
ESS hypothesis. Note that the $\xi_p$ exponents of the three 
simulations lie within the error bars, and the three simulations 
show departures from the self-similar K41 or IK scaling. For values 
of $p$ larger than 6, effects associated with the limited amount of 
spatial statistics can be observed in all the runs.

\section{Discussion and outlook}\label{discuss-sec}

Sufficient resolution for studying directly high Reynolds number flows as
encountered in geophysics and astrophysics is today well 
beyond technological limits. Closures such as the Lagrangian-averaged 
alpha-model can reduce the computational burden by reducing the 
resolution requirements. However, to be used as a model of hydrodynamic 
or magnetohydrodynamic turbulence, or for applications in astrophysics 
and geophysics, detailed knowledge of the ability of the \lansa or \lamhda
equations to capture key features of turbulent flows is required.

The \lansa and \lamhda equations have been tested against direct numerical 
simulations in a variety of problems (see \eg \cite{CFHOTW99CHMZ99} for 
neutral fluid studies and \cite{lamhd2d,lamhd3d,prldynamo} for studies 
in conducting fluids). Most of these works compared the time evolution 
of ideal invariants for forced and free decaying turbulence, as well as 
the evolution of energy spectra. Also, some statistical comparisons were 
performed (\eg studying the behavior of probability density functions). 
In this work, we apply a more stringent test to these models. Intermittency 
is a well known feature of turbulent flows, associated with the existence 
of strong events localized both in space and time. Intermittency can 
trigger large scale events, affect the transport coefficients, or 
give rise to corrections in the turbulent scaling. As a result, whether 
a model can capture the statistics of intermittent events is of utmost 
importance to model astrophysical or geophysical flows. The study of 
intermittency also requires computation of high order statistics, 
thereby extending previous comparisons between DNS and $\alpha$-models.

An extension of the K\'arm\'an-Howarth theorem (KH-$\alpha$) was proven 
for the \lansa equations in Ref. \cite{Ho2002c}. As a corollary of this 
theorem, Kolmogorov four-fifths law and Kolmogorov's energy spectrum can 
be derived for the \lansa equations at scales larger than $\alpha$. In this 
work, we extended the KH-$\alpha$ theorem to the \lamhda case, proving as 
a result that for scales larger than $\alpha$ the \lamhda equations satisfy 
the scaling laws previously known for MHD turbulence \cite{PoPu1998,Ch1951}. 
This is an important result, since MHD turbulence involves two coupled fields 
(the velocity and magnetic fields) and can display different power laws in 
the inertial range according to the regime of interest. While LES often 
impose a particular regime and a power law \cite{LESMHD,AMK+01,TFS94,KM04},
the \lamhda 
equations are shown to satisfy the general scalings satisfied by the MHD 
equations without any hypothesis about the scaling followed in the inertial 
range.

The extension of the KH-$\alpha$ theorem to the \lamhda case also allows 
us to define correlation and structure functions in the $\alpha$-model. 
With these functions, the analysis of anomalous scaling and intermittency 
can be performed. Numerical simulations were carried out both for freely 
decaying and for forced two dimensional MHD turbulence, solving directly 
the MHD equations, and employing the \lamhda equations at 1/2 and 1/4 
resolution (a case equivalent to 1/8 of the DNS resolution was also 
considered for forced turbulence). In the forced runs, we have averaged 
statistics over 189 turn-over times (and up to $\sim2\cdot10^8$ points) 
to test if the \lamhda equations reproduces intermittent turbulent 
behavior. The scaling of the third-order structure function was tested
and linear scaling with length (down to length $\alpha$) was observed, in 
good agreement with corollaries of the extended KH-$\alpha$ theorem and 
exact laws in MHD turbulence \cite{PoPu1998}. The \lamhda equations also 
capture high-order statistics (up to and including order 8) and the 
anomalous scaling of the longitudinal structure function exponents, with 
a net gain in speed close to a factor of 16.  \neu{}{For lower order
structure functions, very little contamination of the scaling could be detected
at scales larger than $\alpha$. On the other hand, for the highest computed 
order, fluctuations in the scaling are observed for the runs with the 
smallest resolution. Note that we would not expect any scaling to
be preserved for $\alpha$ so large that no inertial range remains.

Turbulence closures are never unique. The present case may owe its
success not to its particular form, but rather to its general properties
that it (1) preserves physical avenues of nonlinear energy exchange and
(2) allows correct vortex stretching. These two properties derive from
its origin via a Lagrangian-averaged Hamilton's principle. The
derivation also identifies the appropriate dissipation for proper
energy decay, which involves an enhanced resistivity, but not an enhanced
viscosity. Together, the Navier-Stokes viscosity and the enhanced
resistivity produce regularization ({\sl e.g.}, existence and uniqueness of
strong solutions and their global attractor of finite Hausdorf
dimension, to be discussed elsewhere). In turn, these choices of
viscosity and resistivity allow the intermittency found here, which
might have otherwise been suppressed.}

Relying on the fact that, contrary to fluids, two dimensional MHD turbulence 
displays a direct cascade of energy and intermittency, we could show that the \lamhda equations reproduce intermittency features of turbulent flows and thus we postulate that  these results will carry over to the three-dimensional case and thus these 
results could be also of relevance to the modeling of neutral fluids.
\neu{}{Future challenges will include implementation of the \lamhda model
in domains with boundaries and the study of intermittency for magnetic
Prandtl numbers besides unity. The choices of boundary conditions may be
expected to strongly influence the solution behavior. Of course, this
matter is beyond the scope of the present article.}

\subsubsection*{Acknowledgements}

Computer time was provided by NCAR. The NSF grant CMG-0327888
at NCAR supported this work and is gratefully acknowledged.
DDH is grateful for partial support by the US Department of Energy, under
contract W-7405-ENG-36 for Los Alamos National Laboratory, and Office of
Science ASCAR/AMS/MICS.


\begin{thebibliography}{}

\bibitem{HOS01}
P. Hoyng,  M. A. J. H. Ossendrijver, and D. Schmitt,
``The geodynamo as a bistable oscillator,''
Geophys. Astrophys. Fluid Dyn. {\bf 94}, 263 (2001);
P. Hoyng, D. Schmitt, and M. A. J. H. Ossendrijver,
``A theoretical analysis of the observed variability of the geomagnetic dipole
field,''
Phys. of Earth and Plan. Int. {\bf 130}, 143-157 (2002).
\bibitem{H93}
P. Hoyng,
``Helicity fluctuations in mean field theory: an explanation for the variability of the solar cycle?,''
Astron. Astrophys. {\bf 272}, 321 (1993);
A. J. H. Ossendrijver, and P. Hoyng,
``Stochastic and nonlinear fluctuations in a mean field dynamo,''
Astron. Astrophys. {\bf 313}, 959-970 (1996).
\bibitem{CBLSJ04}
P. Charbonneau,
``Multiperiodicity, Chaos, and Intermittency in a Reduced Model of the Solar Cycle,''
Sol. Phys. {\bf 199}, 385-404 (2001);
P. D. Mininni, D. O. G\'omez, and G. B. Mindlin,
``Biorthogonal Decomposition Techniques Unveil the Nature of the Irregularities Observed in the Solar Cycle,''
Phys. Rev. Lett. {\bf  89}, 061101 (2002);
P. Charbonneau, G. Blais-Laurier, and C. St-Jean,
``Intermittency and Phase Persistence in a Babcock-Leighton Model of the Solar Cycle,''
Astrophys. J. {\bf 616}, L183-L186 (2004);
P. D. Mininni, and D. O. G\'omez,
``A new technique for comparing solar dynamo models and observations,''
Astron. Astrophys. {\bf 426}, 1065-1073 (2004).
\bibitem{KSM99}
J. R. Kulkarni, L. K. Sadani, and B. S. Murthy,
``Wavelet Analysis of Intermittent Turbulent Transport in the Atmospheric Surface Layer over a Monsoon Trough Region,''
Boundary-Layer Meteorology {\bf 90}, 217 - 239 (1999).
\bibitem{CM98}
S. Cerutti and C. Meneveau,
``Intermittency and relative scaling of subgrid-scale energy dissipation 
in isotropic turbulence,'' 
Phys. Fluids {\bf 10}, 928-937 (1998).
\bibitem{meneveaukatz}
C. Meneveau and J. Katz, 
``Scale-Invariance and turbulence models for Large-Eddy Simulation,'' 
Annu. Rev. Fluid Mech. {\bf 32}, 1-32 (2000).
\bibitem{LESMHD}
A. Pouquet, U. Frisch, and J.  L\'eorat, 
``Strong MHD helical turbulence and nonlinear dynamo effect,'' 
J. Fluid Mech. {\bf 77}, 321-354 (1976); 
A. Yoshizawa, 
``Subgrid modeling for magnetohydrodynamic turbulent shear flows,'' 
Phys. Fluids {\bf 30}, 1089-1095 (1987). 
\bibitem{AMK+01}
D. W. Longcope and R. N. Sudan,
``Renormalization group analysis of reduced magnetohydrodynamics with
application to subgrid modeling,''
Phys. Fluids B {\bf 3}, 1945-1962 (1991);
O. Agullo, W.-C. M\"uller, B. Knaepen, and D. Carati, 
``Large eddy simulation of decaying magnetohydrodynamic turbulence with
dynamic subgrid-modeling,'' 
Phys. Plasmas {\bf 8}, 3502-3505 (2001); 
W.-C. M\"uller and D. Carati, 
``Dynamic gradient-diffusion subgrid models for incompressible 
magnetohydrodynamic turbulence,'' 
Phys. Plasmas {\bf 9}, 824-834 (2002).
\bibitem{TFS94}
M. Theobald, P. A. Fox, and S. Sofia, 
``A subgrid-scale resistivity for magnetohydrodynamics,'' 
Phys. Plasmas {\bf 1}, 3016-3032 (1994);
Y. Zhou, O. Schilling, and S. Ghosh,
``Subgrid scale and backscatter model for magnetohydrodynamic turbulence based on closure theory: Theoretical formulation,''
Phys. Rev. E {\bf 66}, 026309 (2002).
\bibitem{KM04}
B. Knaepen and P. Moin, 
``Large-eddy simulation of conductive flows at low magnetic Reynolds number,'' 
Phys. Fluids {\bf 16}, 1255-1261 (2004);
Y. Ponty, H. Politano, and J.-F. Pinton,
``Simulation of Induction at Low Magnetic Prandtl Number,''
Phys. Rev. Lett. {\bf 92}, 144503 (2204).
\bibitem{lamhd2d}
P.D. Mininni, D.C. Montgomery, and A. Pouquet, 
``A numerical study of the alpha model for two-dimensional 
magnetohydrodynamic turbulent flows,'' 
Phys. Fluids {\bf 17}, 035112 (2005).
\bibitem{lamhd3d}
P.D. Mininni, D.C. Montgomery, and A. Pouquet, 
``Numerical solutions of the three-dimensional magnetohydrodynamic 
alpha-model,'' 
Phys. Rev. E {\bf 71}, 046304 (2005).
\bibitem{prldynamo} 
Y. Ponty, P.D. Mininni, D.C. Montgomery, J.-F. Pinton, H. Politano, and 
A. Pouquet, 
``Numerical study of dynamo action at low magnetic Prandtl numbers,'' 
Phys. Rev. Lett. {\bf 94}, 164502 (2005).
\bibitem{Ho2002}
D.D. Holm,
``Fluctuation effects on 3D Lagrangian mean and Eulerian mean fluid motion,'' 
Physica D {\bf 133}, 215-269 (1999);
D.D. Holm,
``Averaged Lagrangians and the mean dynamical effects of fluctuations 
in continuum mechanics,'' 
Physica D {\bf 170}, 253-286 (2002);
D.D. Holm,
``Lagrangian averages, averaged Lagrangians, and the mean effects 
of fluctuations in fluid dynamics,'' 
Chaos {\bf 12}, 518-530 (2002).
\bibitem{HoMaRa1998b} 
D.D. Holm, J.E. Marsden, and T.S. Ratiu,
``The Euler--Poincar\'e equations and semidirect products with 
applications to continuum theories,'' 
Adv. Math. {\bf 137}, 1-81 (1998).
\bibitem{montgo_02}
D. Montgomery and A. Pouquet, 
``An alternative interpretation for the Holm `alpha model',''
Phys. Fluids {\bf 14}, 3365--3366 (2002).
\bibitem{biskamp}
D. Biskamp and E. Schwarz, 
``On two-dimensional magnetohydrodynamic turbulence,'' 
Phys. Plasmas {\bf 8}, 3282-3292 (2001).
\bibitem{K41b}
A.N. Kolmogorov, 
``The local structure of turbulence in incompressible viscous fluid 
for very large Reynolds number,'' 
Dok. Akad. Nauk SSSR {\bf 30}, 9-13 (1941); 
``On degeneration (decay) of isotropic turbulence in an incompressible 
viscous liquid,'' 
Dok. Akad. Nauk SSSR {\bf 31}, 538-540 (1941); 
``Dissipation of energy in locally isotropic turbulence,'' 
Dok. Akad. Nauk SSSR {\bf 32}, 16-18 (1941).
\bibitem{iroshnikovRHK}
P.S. Iroshnikov, 
``Turbulence of a conducting fluid in a strong magnetic field,'' 
Sov. Astron. {\bf 7}, 566-571 (1963);
R.H. Kraichnan, 
``Inertial-range spectrum of hydromagnetic turbulence,'' 
Phys. Fluids {\bf 8}, 1385-1387 (1965).
\bibitem{galtier2000} S. Galtier, S.V. Nazarenko, A.C. Newell, A. Pouquet, 
``A weak turbulence theory for incompressible magnetohydrodynamics,'' 
J. Plasma Phys. {\bf 63}, 447-488 (2000).
\bibitem{PoPu1995} 
H. Politano and A. Pouquet, 
``Model of intermittency in magnetohydrodynamic turbulence,'' 
Phys. Rev. E {\bf 52}, 636-641 (1995).
\bibitem{PoPu1998} 
H. Politano and A. Pouquet, 
``von K\'arm\'an-Howarth equation for magnetohydrodynamics and its 
consequences on third-order longitudinal structure and correlation 
functions,'' 
Phys. Rev. E {\bf 57}, 21-24 (1998); 
``Dynamical length scales for turbulent magnetized flows,'' 
Geophys. Res. Lett. {\bf 25}, 273-276 (1998).
\bibitem{FoHoTi2001} C. Foias, D.D. Holm, and E.S. Titi, 
``The Navier-Stokes-alpha model of fluid turbulence,'' 
Physica D {\bf 152}, 505-519 (2001).
\bibitem{KaHo1938}
T. von K\'arm\'an and L. Howarth,
``On the statistical theory of isotropic turbulence,'' 
Proc. Roy. Soc. London, Ser. A {\bf 164}, 192-215 (1938).
\bibitem{Ch1951}
S. Chandrasekhar, 
``The invariant theory of isotropic turbulence in magneto-hydrodynamics,'' 
Proc. Roy. Soc. London, Ser. A {\bf 204}, 435-449 (1951).
\bibitem{Ho2002c}
D.D. Holm, 
``K\'arm\'an-Howarth theorem for the Lagrangian-averaged 
Navier-Stokes-alpha model of turbulence,'' 
J. Fluid Mech. {\bf 467}, 205-214 (2002).
\bibitem{Ch1950}
S. Chandrasekhar, 
``The theory of axisymmetric turbulence,'' 
Philos. Trans. R. Soc. London, Ser. A {\bf 242}, 557-577 (1950).
\bibitem{Fr1995} U. Frisch, 
{\it Turbulence: the legacy of A.N. Kolmogorov} 
(Cambridge University Press, Cambridge, 1995).
\bibitem{Ey1996} G.L. Eyink,
     ``Exact results on stationary turbulence in
     2D: consequences of vorticity conservation,'' 
     Physica D \textbf{91}, 97--142 (1996).
\bibitem{CoETi1994Ey2002} P. Constantin, W. E, and E.S. Titi, 
``Onsager's conjecture on the energy conservation for solutions of Euler's,'' 
Commun. Math. Phys. {\bf 165}, 207-209 (1994);
G.L. Eyink,
``Local 4/5-law and energy dissipation anomaly in turbulence,'' 
Nonlinearity {\bf 16}, 137-145 (2003).
\bibitem{Alex} A. Alexakis, P.D. Mininni, and A. Pouquet, 
``Shell to shell energy transfer in MHD, Part I: steady state turbulence,'' 
Phys. Rev. E {\bf 72}, 046301 (2005).
P.D. Mininni, A. Alexakis, and A. Pouquet, 
``Shell to shell energy transfer in MHD, Part II: kinematic dynamo,'' 
Phys. Rev. E {\bf 72}, 046302 (2005).
\bibitem{BCB+93}
R. Benzi, S. Ciliberto, C. Baudet, G. Ruiz Chavarria, and R. Tripiccione, 
``Extended self-similarity in the dissipation range of fully developed 
turbulence,'' 
Europhysics Letters {\bf 24}, 275-279 (1993);
R. Benzi, S. Ciliberto, R. Tripiccione, C. Baudet, F. Massaioli, and S. Succi,
``Extended self-similarity in turbulent flows,'' 
Phys. Rev. E {\bf 48}, R29-R32 (1993);
R. Benzi, L. Biferale, S. Ciliberto, M.V. Struglia, and R. Tripiccione,
``Generalized scaling in fully developed turbulence,'' 
Physica D {\bf 96}, 162-181 (1996).
\bibitem{HoMaRa1998a}
D.D. Holm, J.E. Marsden, and T.S. Ratiu,
``Euler-Poincar{\' e} Models of Ideal Fluids with Nonlinear Dispersion,'' 
Phys. Rev. Lett. {\bf 80}, 4173-4176 (1998).
\bibitem{HB04}
N. E. L. Haugen and A. Brandenburg,
``Inertial range scaling in numerical turbulence with hyperviscosity,''
Phys. Rev. E {\bf 70}, 026405 (2004).
\bibitem{She94} Z.-S. She and E. L\'ev\^eque, 
``Universal scaling laws in fully developed turbulence,'' 
Phys. Rev. Lett. {\bf 72}, 336-339 (1994).
\bibitem{PPC98}
H. Politano, A. Pouquet, and V. Carbone,
``Determination of anomalous exponents of structure functions in 
two-dimensional magnetohydrodynamic turbulence,'' 
Europhysics Letters {\bf 43}, 516-521 (1998).
\bibitem{GoPoPo1999}
T. Gomez, H. Politano, and A. Pouquet, 
``On the validity of a nonlocal approach for MHD turbulence,'' 
Phys.  Fluids {\bf 11}, 2298-2306 (1999).
\bibitem{CFHOTW99CHMZ99}
S.Y. Chen, C. Foias, D.D. Holm, E.J. Olson, E.S. Titi, and S. Wynne, 
``A connection between the Camassa-Holm equations and turbulence in 
pipes and channels,'' 
Phys.  Fluids  {\bf 11}, 2343-2353 (1999);
S.Y. Chen, D.D. Holm, L.G. Margolin, and R. Zhang,
``Direct numerical simulations of the Navier-Stokes alpha model,'' 
Physica D {\bf 133}, 66-83 (1999).
\bibitem{HB04}
N. E. L. Haugen and A. Brandenburg,
``Intertial range scaling in numerical turbulence with hyperviscosity,''
Phys. Rev. E {\bf 70}, 026405 (2004).
\end{thebibliography}
\end{document}